\documentclass[prd,tightenlines,twocolumn]{revtex4}
\usepackage{amsmath}
\usepackage{amssymb}
\usepackage{graphicx}
\usepackage{bm}
\usepackage{enumerate}
\usepackage{color}

\newcommand{\be}{\begin{equation}}
\newcommand{\ee}{\end{equation}}
\newcommand{\la}{\langle}
\newcommand{\ra}{\rangle}
\newcommand{\bea}{\begin{eqnarray}}
\newcommand{\eea}{\end{eqnarray}}
\newcommand{\ba}{\begin{eqnarray}}
\newcommand{\ea}{\end{eqnarray}}

\begin{document}

\title{ QCD Topology at Finite Temperature: \\ Statistical Mechanics of Selfdual Dyons  }

\author{ 
Pietro Faccioli$^{1,2}$ and Edward Shuryak$^3$}
\affiliation{$^1$ Physics Department,\\ Trento University,\\ Via Sommarive 14, Povo (Trento), I-38100,  Italy}
\affiliation{$^2$ INFN, Gruppo Collegato di Trento,\\ Via Sommarive 14, Povo (Trento), I-38100,  Italy}
\affiliation{$^3$Department of Physics and Astronomy, \\ Stony Brook University,\\
Stony Brook, NY 11794, USA}

\date{\today}

\begin{abstract} Topological phenomena in gauge theories have long been recognized as the driving force for chiral symmetry breaking and confinement. These phenomena can be conveniently investigated in the semi-classical picture, in which the topological charge is entirely carried by (anti-)self-dual gauge configurations. In such an approach,  it has been shown that near the critical temperature, the non-zero expectation value of the Polyakov loop (holonomy)  triggers the ``Higgsing" of the color group, generating the splitting of instantons into $N_c$ self-dual dyons.
 A number of lattice simulations have provided some evidence for such dyons, and traced their relation with specific observables, such as the Dirac eigenvalue spectrum.  In this work,  we formulate a model, based on one-loop
 partition function and including Coulomb interaction, screening and fermion zee modes. We then perform the first numerical
 Monte Carlo simulations of a statistical ensemble of self-dual dyons,
 as a function of their density, quark mass and the number of flavors.
  We study different dyonic two-point correlation functions and 
 we compute the Dirac spectrum, as a function of the ensemble diluteness and the number of quark flavors.
 \end{abstract}

\maketitle
\section{Introduction}
Topological phenomena in gauge theories  have been discovered more than three decades ago, and
remain the subject of intense theoretical research  ever since. In particular, magnetic objects (monopoles)
have been identified as a possible source of confinement \cite{Mandelstam:1974pi,'tHooft:1981ht}, while instantons have been proposed as the 
driving mechanism for chiral symmetry breaking \cite{Shuryak:1981ff,Diakonov:1985eg}.

 The index theorem establishes a direct connection between the vacuum topology, and zero-eigenvalue solutions of the Dirac equation, i.e. the so-called fermionic zero-modes. These quark states are insensitive to any perturbative fluctuation of the gauge field, hence encode purely non-perturbative QCD dynamics.
 Furthermore,
 lattice simulations have shown that the Dirac eigenstates with near-zero eigenvalues --- also known as the "zero-mode zone" (ZMZ)--- directly correlate with  \emph{local} fluctuations of the topological charge density.
 After  filtering out quantum fluctuations, lattice fields reveal
  nearly (anti) self-dual smooth fields responsible for topology and ZMZ states~\cite{Gattringer}.
Using only fermionic states attributed to the ZMZ  (a  tiny
 subset of Dirac eigenstates, of only about $\sim 10^{-4}$ of all eigenstates) one finds
 the correct pion mass, quark condensate as well as many other hadronic properties.
On the other hand, filtering out the ZMZ states removes the chiral symmetry breaking and  leads to drastic 
 changes in the  hadronic spectrum computed on the lattice. In particular,  some masses get shifted by as much as $\sim 30\%$ and 
  parity-doublets appear (for a  recent analysis, see e.g.  \cite{Glozman:2012hw}). 
  
  This body of results coherently support a picture in which the non-perturbative chiral dynamics in vacuum is mediated by instantons. 
  Indeed, instanton model calculations (for a review see \cite{Schafer:1996wv}), have been very successful in reproducing  the mass and electro-magnetic structure of pions~\cite{pionFF}, vector mesons~\cite{hadronicresonances},  nucleons~\cite{Negele, diquarks, masses, nucleonFF1, nucleonFF2, nucleonFF3} and even the $\Delta I = 1/2$ rule for hyperon \cite{delta12} and kaon \cite{delta12kaon} non-leptonic decays. 
    
In the instanton picture, the width of the ZMZ depends on the size of 
the typical ``hopping" matrix element of the Dirac operator between two instantons, which is 
of the order  $\sim \bar \rho^2/\bar R^3\sim$~20~MeV, where $\bar \rho$ is the typical instanton size and $\bar R$ the typical inter-instanton density \cite{Shuryak:1981ff}. This value is comparable to the typical light quark masses used
in may lattice simulations, and this explains why  the corresponding results display significant deviations from the  naive chiral perturbation theory predictions. Furthermore, the specific shape of the density of eigenvalues $\rho(\lambda)$ in the ZMZ depends crucially on the theory parameters, such as the number of light fermions $N_f$.
 
%
   
    In this work, we will further investigate topological phenomena in the semi-classical picture, focusing on temperatures close to those at which  the expectation value of the Polyakov line
   \be \label{holonomy}
\langle  P({\bf x}) \rangle = \langle \exp\left(~i \int_0^\beta
d x_4 ~A_4^a({\bf x},x_4) ~\frac{\lambda^a}{2}~\right) \rangle 
\ee
drastically changes from 1 to 0. 

The gauge invariant expectation value (\ref{holonomy})  defines the holonomy  of the gauge connection corresponding to a full circle around the periodic time direction and is related to the free-energy $F_q$ of a single static quark:
 \be
 \la P\ra \sim \exp\left(-F_q/T\right).
 \ee
Hence, a drastic suppression of $ \la P\ra $  reflects the onset of the confinement phenomenon and we shall denote the corresponding critical temperature with $T_c$.

From Eq. (\ref{holonomy}) it follows that a non-trivial holonomy (i.e. $\langle  P({\bf x}) \rangle \ne 1$) 
reflects a non-vanishing  vacuum expectation value (VEV)  
of the $A_4({\bf x}, x_4)$ component of the gauge field.
In the semiclassical picture, this condition is not fulfilled by the standard instantons and calorons (finite temperature instantons). Indeed, in such topological classical solutions, the $A_4$  component of the gauge field vanishes at spatial infinity,  $\lim_{|{\bf x}|\to \infty} A_4({\bf x}, x_4)  =0$.

Classical gauge configurations, generalizing the  instanton solution to a non-trivial holonomy case, 
are called the Kraan-van-Baal-Lee-Lu (KvBLL) calorons \cite{Lee:1997vp,Kraan:1998sn}. 
 These authors have shown that instantons basically split into  certain substructures, the  $N_c$ ``constituent dyons".   
  
    The  $4 N_c$ collective variables for the constituent dyons are different from that of standard instantons:
   while each instanton is identified its  position, size and color orientation, a dyon  is specified  only  by its position and its  Abelian phases. 
   The solution and the measure for the KvBLL configurations  correctly returns to the standard instanton one in the limit of vanishing holonomy.  
 For review of instanton-dyons see   \cite{Bruckmann:2003yq}, \cite{Diakonov:2009ln}  and references therein. 
 
In general case, the  holonomy can be parametrized in terms the VEV of $A_4$ at spatial infinity,
according to the usual Cartan subalgebra notation. 
Such an asymptotic $A_4$ field can be taken to be spatially constant and diagonal in color space, 
\be 
A_4(\infty)=(2\pi T) ~\textrm{diag}(\mu_1... \mu_{Nc}); \,\, \,\,\, \sum_i \mu_i=0
\ee
The latter condition follows from imposing a null trace.

Inserting such a constant field plus and its quantum fluctuations into the gauge action, one finds 
that the Yang-Mills  commutator 
 generates mass terms for all non-diagonal components of the gluon field.
On the other hand, diagonal components of the quantum gauge field commute with the holonomy and remain massless.
 
It is convenient to adopt a cyclic-symmetric notations such that
  $\mu_{Nc+1}=1+\mu_1$ and introduce the differences between subsequent holonomy parameters,
\be  
\nu_i=\mu_{i+1}-\mu_{i}  ; \,\, \,\,\, \sum_i \nu_i=1. 
\ee
These differences determine the masses of the non-diagonal $(i+1,i)$ 
components of the gauge field.  Since the dyon cores
are made of such field components,   the  i-th dyon has the core  size  $\sim 1/(2\pi T \nu_i)$.
On the other hand, the long-range dyon interaction is due to massless diagonal fields and is governed by their
electric and magnetic charges (see e.g. Ref.~\cite{Diakonov:2009ln} ).

\begin{figure}[t] 
   \centering
   \includegraphics[width=8cm]{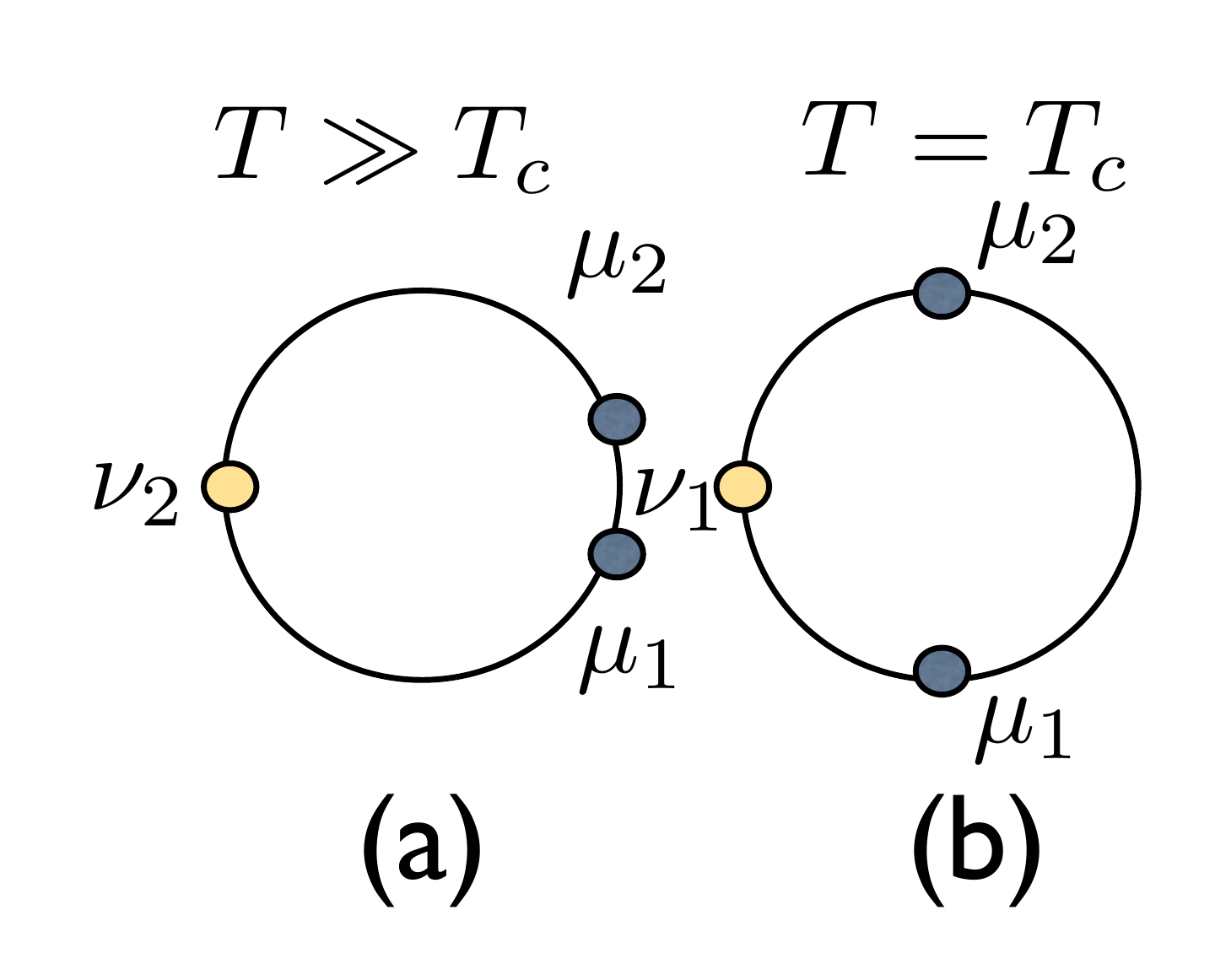} 
   \caption{Location of the holonomy eigenvalues $\mu_1,\mu_2$ shown by (blue) darker small circles,
    for high temperatures (a) and for $T=T_c$ (b). Their differences $\nu_1=\mu_2-\mu_1, \nu_2=\mu_3-\mu_2$
    are also indicated. The (yellow) lighter circles on the left at the phase $\pi$ indicate the case of physical anti periodic
    fermions. 
   \label{fig:holonomy}
   }
\end{figure}

In this paper, we restrict our attention to  the 
 SU(2) gauge group. In this case,  the holonomy parameterization simplifies,
 since there is only one diagonal generator $\tau^3$ and a  single holonomy parameter,   $\mu= \mu_1=-\mu_2$.
 In Fig.\ref{fig:holonomy} we show the locations of the holonomy eigenvalues at two different temperatures. 
 At high $T$ (case (a))  the holonomy is close to a zero phase, so $\nu_2=2~\mu$ is small. Respectively, the mass of the first dyon, called M,
 is also small. The mass of the second dyon is proportional to the complementary part of the phase circle $\nu_2$, which at high $T$ takes nearly the whole circle.   The corresponding $L$ dyon is thus heavy. 
 
 As $T\rightarrow T_c$, the holonomy moves toward $\pm 1/4$ of the circle, as shown in
 case (b): and in this case both dyons have the same mass. This position corresponds to phases $\mu_i=\pm 1/4$ and the
 Polyakov line vanishes $P\rightarrow  \cos(\pi/2)=0$, which physically corresponds to the confinement of heavy charges (quarks).  
 
  In Table I we list the quantum numbers of the possible dyon solutions for the SU(2) gauge group. In such a gauge theory there are 4 distinct dyons, denoted with $M$, $\bar M$, $L$ and $\bar L$, which are characterized by different (color-) magnetic and electric charges.  For the general $\text{SU}(\text{N}_c)$ case  there are $M_1,M_2...M_{Nc}$ so-called ``static" dyons with all diagonal charges and just one so-called ``twisted" dyon $L$.

From the technical point of view,  the  statistical physics of the dyon ensemble is quite challenging compared to that of 
the ``instanton liquid"  \cite{Schafer:1996wv}. This is mostly
 because, on top of topological phenomena induced by the  zero-modes of light fermions,
 one has to deal with long-range Coulomb forces
and even linearly growing potentials due to a ``screening" effect, as will be discussed below.


A first, qualitative     
discussion of dyonic ensemble has been performed  by T.~Sulejmanpasic and one of us \cite{Shuryak:2012aa}.
It was pointed out that a dyonic system can realize several distinct  phases, ranging from a gas of individual $\bar{L} L$ clusters at high $T$, to a strongly-coupled liquid plasma at $T\approx T_c$.
 Ref. \cite{Shuryak:2012aa} also includes an extensive
discussion of the lattice-based phenomenology related to self-dual dyons. Extension of that work to the case of adjoint fermions will be given in \cite{Tin2}. 

The present paper can be considered as a {\em quantitative} continuation of that study,  we shall define a partition function of the dyonic ensemble
which is amenable to practical 
many-body computer simulations, perform the first numerical simulations and report on the results.

\begin{table}[t!]
\begin{tabular}{| c | c|c| c| c|} \hline
name & E & M & mass \\ \hline
$M$         & + &+ &$v$ \\
$\bar{M}$ & + &- &$v$ \\
$L$          & - & - &$2\pi T -v$ \\
$\bar{L}$ & - & + &$2\pi T -v $ \\ \hline
\end{tabular} 
\caption{ The electric and magnetic charges and the mass (in units of $8\pi^2/g^2 T$) for the 4 different kinds of SU(2) dyons.} \label{tab_su2dyons}
\end{table}

The partition function of the dyon ensemble can be schematically written as 
\be\label{Zdef}
 Z = \int \{dX_i \}  e^{-S_{c}}~{\textrm{det}~G }~ \textrm{det} F_{zm}~{ \textrm{det}' F_{nzm}\over \sqrt{\textrm{det}'B}}  
  \ee
 where  the first factor is the product of the differentials of all the collective variables, the exponent contains the classical actions of all dyons, while the two subsequent determinants of matrices $G$ and $ F_{zm}$ are
 related  to the bosonic and fermionic zero-modes, respectively. The two remaining  (primed) determinants correspond to the nonzero-modes of the one-loop 
 bosonic and fermionic operators, describing the small linearized perturbations of the classical solution.
 (As  usual, the latter determinants are divergent and need to be regularized, e.g. in the Pauli-Villars scheme, so that 
  the  UV bare coupling $g$ in the action is combined with the UV cut-off into a physical renormalized charge.) 
A quantitative characterization of the dyonic ensemble  can be achieve  by performing Monte Carlo sampling of the partition function (\ref{Zdef}).

 Before we come to such simulations, let us outline the physical phenomena to be addressed. 
As the temperature drops from $T\sim 3T_c$ to $T_c$, the value of the holonomy changes from zero
to its ``confining value" $v=\pi T$. In the former case the action of the $M$-dyons is small and of the $L$ ones is large,
while at $T<T_c$ their actions become nearly equal. 
At high $T$ ``heavy" $L$ dyons can be bound with antidyons $\bar{L}$
by the fermion exchanges into certain $L \bar{L}$ clusters \cite{Shuryak:2012aa},  with zero magnetic but double electric Abelian charge.
The formation of such clusters are reflected in the double-bump shape of the Dirac eigenvalue spectrum
and a gap around zero: in the gas of clusters the chiral symmetry remain unbroken. As $T$ decreases, the action per dyon
decreases and their density grow. Also the mass of the fermions decreases.  Eventually, at some $T=T_c$, the density is high enough to bridge the gap
in the Dirac eigenvalue spectrum and produce finite density of eigenvalues at zero. This means that the chiral symmetry gets spontaneously broken.

Let us now briefly outline the structure of the paper. The next section describes bosonic interactions of the dyons, which are
separated into approximations made to describe the so called {\em moduli space metric} associated with bosonic zero-modes,
and {\em   electric screening} associated with the nonzero bosonic modes. The next section describes our approximation for the third
essential ingredient of the partition function, the {\em fermionic determinant}.

 The subsequent two sections provide further technical details about the partition function (\ref{Zdef}). Some introduction about
 the instanton dyons and the setting used in our simulations is reported in section \ref{sec_setting}.
The results of the simulations are presented and discussed in section \ref{sec_results}, while the main conclusions are summarized in \ref{sec_summary}. 

\section{Bosonic dyon interactions}
\subsection{Moduli space metric}
In the selfdual (or anti-selfdual) subsectors, the dyons are so-called Bogomolny-Prasad-Sommerfeld
(BPS) protected objects, which means that their total action is completely determined by the total topological charge.
Thus, at the level of classical solutions of the Yang-Mills equations, the total action is independent of the  dyon collective coordinates. 
The corresponding degenerate manyfold of all  classical solutions with a particular topological charge $Q$, known as the ``moduli space",
has certain  metric, which can be calculated from the bosonic zero-modes
of the classical solution in a standard way. The resulting geometry is not simple: for monopoles it was derived in classic works of Atiyah and Hitchin~\cite{AH}. Even a configurations consisting of two monopoles/dyons gives raise to a quite nontrivial manifold, with a specific singularity known as ``bolt" (or arrowhead). Many-body  
moduli are extremely complicated, and to our knowledge their exact metric and geometry has not been explored in any detail.

Obviously, one needs to make some approximations, in order to perform practical simulations.
 Following
Gibbons and Manton \cite{GM} and Diakonov \cite{Diakonov:2009ln}, 
 the invariant volume element for moduli metric can be approximated by the  determinant of a certain matrix $G$
 which is defined in order to incorporate the known limiting cases.
\be
\sqrt{\det \hat g} \approx \det \hat G
\ee

The Jacobian determinant of a single L-M dyon pair has been calculated 
by Diakonov, Gromov, Petrov and Slizovskiy (DGPS) \cite{Diakonov:2004jn}. 
The nonzero-modes lead to the so called screening phenomenon, which we discuss in section \ref{sec_screening}.
For the SU(2) gauge group,  the  matrix $\hat G$ is
\be
\hat G= 
\left(
\begin{array}{c c}
4 \pi \nu_L + \frac{1}{T r_{LM}} & - \frac{1}{T r_{LM}}\\
- \frac{1}{T r_{LM}} & 4 \pi \nu_M +\frac{1}{T r_{LM}} 
\end{array}
\right),
\label{Gdet}
\ee
where the matrix raws and columns correspond to the two dyon types  $L,M$  and $r_{LM}\equiv |x_L - x_M|$. 
We emphasize that the  expression (\ref{Gdet})
turns out to be correct at all distances, not only for asymptotically large separations.  It was
also shown that in the limit of trivial holonomy (i.e. for $\mu_m\to 0$) or 
at  vanishing temperature, the resulting measure 
 with (\ref{Gdet}) 
reduces to the standard 't Hooft single-instanton measure. 

For sake of completeness we report also the generalization of Eq. (\ref{Gdet}) to an arbitrary $\text{SU(N}_c\text{)}$ group,
\ba
[\hat G]_{m n}&=&  \delta_{m n} (4\pi \nu_m + \frac{1}{T|x_m - x_{m-1}|} + \nonumber\\ &&
 \frac{1}{T|x_m - x_{m+1}|} )  - \frac{\delta_{m n-1}}{T|x_m - x_{m+1}  |}  \nonumber\\ &&
  - \frac{\delta_{m n+1}}{T|x_m - x_{m-1}  |}, 
\ea
where the indexes $m,n$ run over the different types of dyons.

The effective potential generated by the {moduli metric} can be defined as usual as
$V_{eff}\equiv-T \log \textrm{det} (G)$. Upon expanding it in powers of $1/r$,   one recovers  the usual Coulombic potential at large distances.

A classical Coulomb gas is known to be unstable: particles can fall onto each other.   Fortunately,  a pair of non-identical dyons is not affected by such a problem. Indeed, the Coulomb term is
 complemented by the higher powers of $1/r$ of opposite signs.  As a result, the total effective potential turns out to be only logarithmically divergent at small distances, thus the integral which appears in the partition function, 
\be 
\int d^3 r \det \hat G = \int d^3 r \left(\frac{1}{r} + \textrm{const.}\right)
\ee
is well convergent. This expression implies that the probability of finding two dyons at the same point vanishes. 


Let us now discuss the case in which there are dyons of the same kind.
 Simplifying the famous Atiyah-Hitchin 
many-monopole metrics \cite{AH},
Gibbons and Manton \cite{GM} have shown that the weight for $K$ identical dyons 
reads:
\be
W = \frac{1}{K!} \int dx_1 \ldots dx_K \det \hat G^\textrm{ident}, 
\ee
where 
\be\label{Gident}
\hat G^\textrm{ident}_{i j} = \left\{
    \begin{array}{c r}
    4 \pi \nu_m - \sum_{k\ne i=1}^K \frac{2}{T|x_i-x_k|}, &i=j\\
     \frac{1}{T|x_i-x_j|} & i\ne j
\end{array}\right.
\ee
 Note  the inversion of the signs here, as compared to the case of non-identical dyons. 
We emphasize the this expression is exact only for large inter-dyon separations, and corrections may be important at small distances.

For a pair of identical dyons, $\textrm{det}G= 4\pi \nu (4\pi \nu- 1/Tr)$,
which remains positive  only for distances larger than some ``core size" $r>1/(4\pi \nu T)$. Note also that  the volume element $\textrm{det} G$
vanishes $linearly$ at the core, which follows from the famous ``bolt" geometry.

Diakonov \cite{Diakonov:2009ln}  nicely combined the metric tensors for
different-kind and same-kind  dyons into
one symmetric-looking metric, describing the moduli space of arbitrary number of dyons. Let their numbers are
$K_1$ dyons of kind 1, $K_2$ dyons of kind 2,$\ldots, K_N$
dyons of kind $N$. The result is a matrix whose dimension
is the total number of dyons $ (K_1 +\ldots+
K_N)\times(K_1+\ldots+K_N)$ and reads:
\ba 
\label{Gtot}
&&[\hat G]_{m,i, n,j} = \delta_{m n} \delta_{i j} \left( 4 \pi \nu_m - 2 \sum_{k\ne i} \frac{1}{T |x_{m, i}- x_{m, k}|} \right.\nonumber\\ 
&&\left.+
  \sum_k \frac{1}{|x_{m, i}- x_{m+1,k}|} + \sum_k \frac{1}{|x_{m, i}- x_{m-1,k}|} \right)  \nonumber\\ 
 &&  +
 2 \frac{\delta_{m, n}}{|x_{m, i}- x_{m,j}|}- \frac{\delta_{m, n-1}}{|x_{m, i}- x_{m+1,j}|}
 \nonumber\\
 &&-\frac{\delta_{m, n-1}}{|x_{m, i}- x_{m-1,j} |}. 
\ea
In this expression $x_{m, i}$ denotes the coordinate of the i$^\textrm{th}$ dyon, of kind $m$. 


Let us now discuss the short-distance behavior.  Now, the effective potential $V_{eff}=\exp(-\log G)$  has multi-particle terms  with all powers
of $1/r$. To make sense of it, it is instructive
to first calculate it for some examples of configurations. For example, we can consider a ``square"
consisting of 2 $L$ and 2 $M$ dyons.  In is found that, in all cases, their combined
effect can be described by a weakening of the Coulomb divergence. 

When the distances between dyons become too small, the Coulombic term becomes unreasonably large,
and it needs to be regulated: after all, a dyon is not a point charge and has a certain size.  
In the present work, we adopt a regulation  based on the substitution
\be  
{2\over r} \rightarrow  {2 \over \sqrt{r^2+ a^2}},
\ee
where we have chosen in the simulation the value of the cutoff parameter $a=\pi/T$, leading to vanishing
measure for two identical dyons, in the limit $r \rightarrow 0$ (to be referred to as  ``soft core" ). This is in contrast with imposing that the measure should vanish  at some finite $r$,
as in the famous Attyah-Hitchin metric (to be referred to as ``hard core").

 The configuration space of a multi-dyon ensemble can be sampled with  a standard Metropolis-based Monte Carlo algorithm. Within such an approach,  we shall not  only enforce the global positivity of the metric determinant, but also 
 we will require each eigenvalue of $\hat G$ to be positive.
 This condition is becoming restrictive at high dyon densities, i.e. when fraction of the total volume filled by dyons is $VT^3 \sim O(1)$, eventually
making the Monte Carlo sampling of a very dense dyon ensemble rather inefficient. 
(This kind of computational difficulty is  well known, e.g. in  simulations of classical liquids with hard repulsive cores.)

\subsection{Nonzero-modes and  electric screening} \label{sec_screening}

The DGPS determinant of  the non-zero-modes \cite{Diakonov:2004jn} for a single  LM dyon pair  
contains the so-called ``confining" term,  proportional to the inter-dyon separation and to the ``screening mass"
\be 
\label{screening}
\log \textrm{det'} B =  r_{LM}~\times{  2 \pi M_D^2 \over  T g^2}, \ee
which displays  the so-called electric Debye mass  $M_D^2= (2/3) g^2 T^2 $, in the SU(2) gluondynamics.  Upon generalizing  to any $N_c$ and number of  fundamental fermions $N_f$
this expression becomes \cite{Shuryak:1977ut}.
\be 
M_D^2= g^2 T^2 (N_c/3+N_f/6). \label{eqn_Debye}
\ee

We now briefly elucidate the origin of the linear potential which appears in Eq. (\ref{screening}).
Let us start with the quartic term $[A_\mu A_\nu]^2$ of the Yang-Mills Lagrangian, and imagine  two of those fields are classical
ones, say  from some dyon solution, and two others belong to thermal
perturbative gluons of the heat-bath. While the latter produce the Debye mass, the former generate
the linear dependence on $r_{LM}$, from the volume average of the squared Coulomb large-distance potentials
  \be
V_{12}\sim \langle (A_4)^2\rangle=  \int d^3 x \left| { 1 \over r_L}-{1 \over r_M} \right| ^2 =4\pi~ r_{LM} \label{eqn_linear}
\ee
Note that we consider here neutral $L-M$ pair (i.e. the splitting of one instanton), as a result of which their Coulomb potential cancels out at large distances from the pair and the volume integral is convergent. Clearly, non-neutral configurations cannot be treated  this way. 

We further note that the form (\ref{screening})  can be obtained directly by the instanton screening term calculated by Pisarski and Yaffe \cite{Pisarski:1980md} by recalling that the instanton size  $\rho$ and the $L-M$ separation are related by the expression
\be \pi\rho^2 T= r_{ML}.  
\label{eqn_rho_r_relation} 
\ee
which relates the ``4-d dipole" of the instanton field to the ``3-d dipole" of the dyon $LM$ pair made of opposite charges. 



Let us now work out the corresponding general  formula for screening potential which holds in the many-body case.
The  sum over all dyonic contributions to $A_4$ can be written as
\be
\langle (A_4)^2\rangle=\int d^3 x \left| \sum_i {Q_i \over r_i} \right|^2
\ee
where now the sum runs over all dyons with $Q_i=\pm 1$ is the charge and $ r_i=|\vec x- \vec z_j|$.
 
 One can write $(A_4)^2$ as a double sum, in which we separate the diagonal terms $i=j$ from all non-diagonal terms.
Unless total neutrality is ensured \be \sum Q_i=0 \ee
this integral is divergent at large distances. If there is an overall neutrality, one can
regulate the sum  term by term, by subtracting the corresponding (r-independent) divergency.
Let us say discuss a non-diagonal term $i\neq j$ which we can then be rewritten as 
\be 
{2 \over r_i r_j} \rightarrow \left({2 \over r_i r_j} -{1 \over r_i^2} -{1 \over r_j^2}\right) = \left(\frac{1}{r_i} -\frac{1}{r_j}\right)^2
 \ee 
reducing it to the 2-body case above (\ref{eqn_linear}). Therefore the total answer is a simple generalization
of the linear potential
\be  \la (A_4)^2\ra = 4\pi \sum_{i>j } Q_i Q_j r_{ij}      \ee
with $Q_i=\pm 1$ being the electric charges.
In contrast to linear potential induced by the confining flux tubes,
in this case there is a sum over {\em all pairs} of dyons.
 The sum
therefore has $N(N-1)/2$ terms for $N$ dyons, of various signs, which together enforce local neutrality of the dyon ensemble,
making $A_4$ as small as possible. 
It does not diverge in infinite volume
because distant dyons may be combined with their counterparts of the opposite charge,
producing dipole fields. On a manifold with the topology of a sphere extra care is needed, as discussed in Appendix \ref{app_Coulomb}.

\section{The fermionic determinant}

In evaluating the fermionic determinant, the Dirac operator is approximated by retaining {only the contribution evaluated on the 
subspace of fermionic zero-modes of the individual pseudo-particles $|\phi_0^j\rangle$:}
\be
\textrm{Det}(i \gamma_\mu D^\mu +  i m) \simeq \textrm{Det}( \hat T +  im),
\ee
where
\be
\label{Tijentry}
T_{i j} = \langle \phi_0^i |i \gamma_\mu D^\mu |\phi_0^j\rangle.
\ee
This scheme was well tested in the framework of the instanton liquid model, where it corresponds to summing up all loop diagrams created by 't Hooft effective 
Lagrangian.  One of the most important lessons from that was existence of rather narrow ``zero-mode zone" (ZMZ) of the Dirac eigenvalues $|\lambda | < \sigma \sim$~20~MeV,
which drive all the effects related to chiral symmetry breaking. The narrowness followed from relative diluteness of the instanton ensemble.

 The dyonic formulation  also generates a narrow ZMZ: in fact, at nonzero holonomy, all light quarks become effectively massive, so the so called ``hopping amplitudes" -- i.e. the non-diagonal matrix elements of the Dirac operator between different dyons -- are exponentially suppressed with distance, 
\be T_{i j}  \sim \exp(-M~r) \label{hop} \ee where $r$
is the inter-dyon distance.
$M$ is the so-called  holonomy mass $M$,  should not be confused with the ordinary (i.e. current) quark mass $m$. In particular,
the former does not break chiral symmetry, since it 
arises from the interaction of the quarks with the $A_4$ component of the gluon field. An estimate of the  magnitude of $M$ for various fermions followed from the zero-mode solution can be found e.g. in  Ref \cite{Shuryak:2012aa}. Antiperiodic fermions in SU(2) have zero modes on $L$ dyons, with the (holonomy-induced chirally symmetric) mass
given by
\be  M=  \pi T-v/2 \ee
proportional to the distance from the fermion boundary condition angle
(  the (yellow) lighter circles on the left at the phase $\pi$ in Fig.\ref{fig:holonomy}) to the (blue) dark circles
indicated the holonomy values. If the fermionic boundary phase is changed, the yellow circle moves and then
one should select the {\em smallest} distance between the two holonomy values.

The most important qualitative trend is the following: as the temperature decreases  from high $T$ to $T\approx T_c$, 
the holonomy changes as indicated in Fig.\ref{fig:holonomy}. As a result,  the effective fermion mass $M$ decreases
and  the ``hopping amplitude" (\ref{hop}) increases. Another qualitative tendency is of course the reduction
 of the effective coupling in the dyonic action, which additionaly increases the dyon density and reduces the inter-dyon distance $r$.  
 Both effects eventually conspire towards breaking the  chiral symmetry.

Some of the (block) entries in Eq. (\ref{Tijentry})  are identically vanishing. In particular, the anti-periodic conditions on fermions imply that zero-mode
exist only for $L$ and $\bar L$ type dyons, see e.g. \cite{Shuryak:2012aa} for explicit solution. Hence,
we can ignore the overlap zones involving $M-$ and $\bar M-$ type dyons.  In addition, the Dirac operator mixes only  modes  with opposite chirality, hence also $T_{L  L}=T_{\bar{L}\bar{L}}= 0$. Hence, the Dirac operator reduces to:
\be
\label{D2}
(i~\gamma_\mu \hat D^\mu + im )= \left(
    \begin{array}{c c}
    im  & T_{L \bar L}\\
    T_{L \bar L}^\dagger & im 
    \end{array}\right) 
\ee
where the  the $i j$ element in $\bar{L}L$ block is given by the (approximate) formula  \cite{Shuryak:2012aa} containing the ``holonomy mass"
\be\label{Tij}
T_{\bar{L}L}^{i j} = c \frac{e^{-M r_{i j}}}{\sqrt{1+ M r_{i j}}}
\ee
Notice that the normalization constant $c$ is irrelevant for Monte Carlo ensemble simulations:
it may only be needed if some physical significance will be ascribed to the quark mass $m $
rather than pushing it to zero. 
%
For $N_f$ massless fermion flavors  the determinant simply reads
\be\label{LLBD}
\det i\hat{D} = |\det T_{L \bar L}|^{2 N_f}.
\ee

\subsection{The case of a single dyon molecule}
Let us first consider the case in which there is only one dyonic "molecule" made by $L$-, $\bar{L}$-, $M$- and $\bar{M}$- type dyons.
The fermionic determinant can be viewed as the second-order one-loop diagram in 't Hooft effective Lagrangian, with two vertices with $2N_f$
fermionic propagators in between. 
From Eq. (\ref{LLBD}) and (\ref{Tij}) we get
\be
\det i\hat{D}  = |T_{L \bar L}|^{2 N_f} = e^{ - 2 N_f M r_{L \bar L}- \log (1+ M r_{L \bar L})}
\ee
Neglecting the logarithmic dependence, the formula for the determinant for 1 molecule $(L, M, \bar L, \bar M)$ reads as en effective potential
\be
\label{twomol}
V=-\log\det i\hat{D} = 2 N_f M~ r_{L \bar L}, 
\ee
where again we have dropped an irrelevant normalization factor. Thus, fermion exchange also creates
a  linear confining potential, but now between dyons and antidyons. Also, note that there is no minus sign and that an additional parameter -- the number
  of flavors $N_f$ -- appears as a factor. Therefore, one expects that fermion exchange to generate the tightly bound $L \bar L$ clusters first 
  described in Ref\cite{Shuryak:2012aa}.

\subsection{The case of two dyon molecules}
{This is the simplest example demonstrating how dyonic clusters undergo mutual repulsion. 
To see this we study the two dyon molecule case, restricting to the L-dyon zone and labeling the pseudo particles }with $L_1$, $L_2$, $\bar L_1$ and $\bar L_2$ we find:
\be
\label{Dtwomol}
i~\hat D = \left(
    \begin{array}{c|c  c c c c}
*& L_1& L_2 & \bar L_1 &\bar L_2 \\
\hline
L_1 & 0 & 0 &   T_{L_1 \bar L_1} & T_{L_1 \bar L_2}\\
L_2 & 0 & 0 &  T_{L_2 \bar L_1} & T_{L_2 \bar L_2}\\
\bar L_1 &  T_{L_1 \bar L_1}^\dagger & T_{L_2 \bar L_1}^\dagger &0 & 0\\
\bar L_2 &  T_{L_1 \bar L_2}^\dagger & T_{L_2 \bar L_2}^\dagger & 0  & 0\\
    \end{array}\right)
\ee
Hence 
\ba
\det i D = - |\det T_{L \bar L}|^{2 N_f} = - |  T_{L_1 \bar L_1} T_{L_2 \bar L_2} \nonumber\\
-
T_{L_1 \bar L_2} T_{L_2 \bar L_1}|^{2 N_f} 
\ea
{from which one immediately sees that} the configuration weight vanishes when the two molecules overlap. In the sext sections we shall
we shall perform numerical simulations which will allow us to investigate the consequences of such a ``Fermi repulsion".

\section{The setting} \label{sec_setting}

\subsection{The setting and parameters}
   Let us start by enumerating parameters of our physics problem, which we will 
   group into global, internal and external ones.
  \begin{itemize}
\item   The  \emph{global parameters} are the number of colors $N_c$ of the theory and the number
   of fundamental fermions $N_f$ of degenerate current mass $m$. In this work $N_c=2$ and $N_f=0,1,2,4$.
   One may in addition consider other types of fermions, e.g. adjoint ones, but we leave such an extension to future work.
   \item The \emph{internal  parameters} are those on which the partition function depends, such as the heat-bath temperature
$T$, and the flavor chemical potentials $\mu_i$. Although the $N_c=2$  theory allows for
finite density Monte Carlo simulations, for now we will only discuss the zero density case, $\mu_i=0$,
using $T$ as the single internal parameter. 

\item The \emph{external parameters} are those which \emph{may} be imposed. For example,
one may  consider modifications of the QCD vacuum in the nonzero (QED) magnetic field: it is now in fact
a quite active research field. Another important example of an external parameter is
the CP-odd angle $\theta$, which however have a sign problem and need reweighing or analytic continuation.
None of it will be discussed in this work.
\end{itemize}

The dyonic model is defined by the partition function defined above
and can be studied for any values of such parameters. After this is done,
the results can be {\em mapped/compared } to lattice data using appropriate values of 
these parameters  as inferred from lattice simulations at the corresponding
temperature. 

 One such input is the value of the holonomy parameter $\mu$ at temperature $T$.
 There are rather accurate lattice data defining it, see Appendix \ref{Dyonlattice1}.

 Other  simulation inputs are  the densities $n_i$ corresponding to dyons of various kind $i$.
We will also use their proper volumes defined by their inverse densities, i.e. $V_i \equiv 1/n_i$. 
 At the time of this writing the lattice studies of the dyons are too few to allow for a systematic mapping,
 of the dyon densities at different temperature. While we do not really use such data, for the orientation of the reader we
 put brief review of the information we find in the Appendix \ref{Dyonlattice2}. 

We will only mention here that since the masses of $L$ and $M$ dyons are different, 
their densities on the lattice are different as well. 
 Note that such disparity between the densities of the $L$ and $M$ dyons produces a nonzero
 density of the Abelian electric (but not magnetic) charge. This is of course not a problem in general,
 as it is compensated 
 by an asymmetry of the densities of the electrically charged gluons.
 
  It would however create a
 technical problem  for our simulation setting, as we  focus entirely on the topological (dyonic) sector.
 Furthermore, one has to use a compact manyfold -- a torus or a sphere --  in which all
 the charges must be compensated and field lines accounted for: hence only totally neutral systems are allowed. One possible
 solution to this problem is simply adding to each charge the {\em homogeneously
 distributed} compensating charge: see Appendix  \ref{app_Coulomb} for details. Another  practical solution 
 we adopt in this work is to ignore the extra $M$ dyons: so we  simulate  an equal number of the $M,L$ dyons.

 \begin{figure*}[t] 
   \centering
   \includegraphics[width=7.3cm]{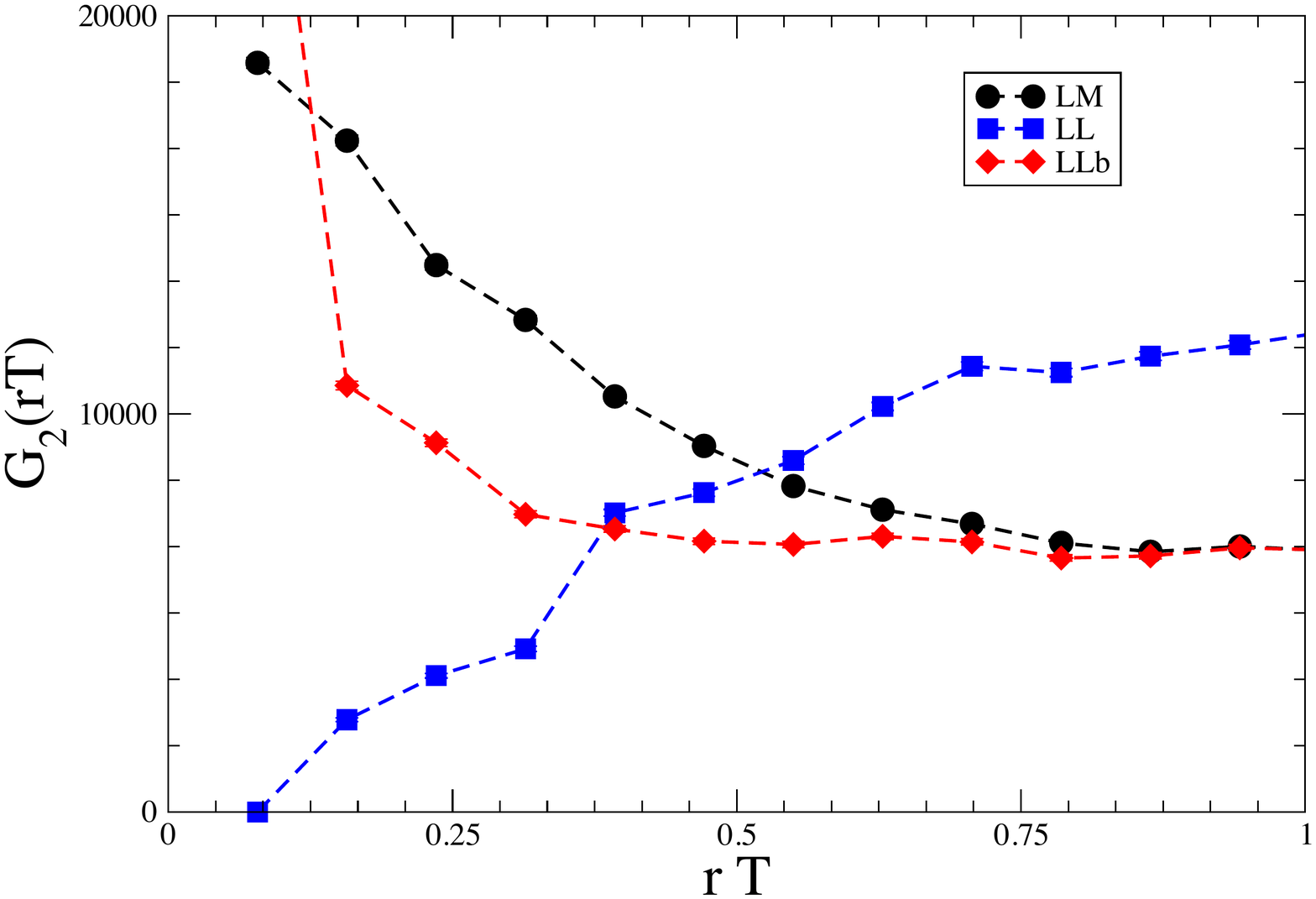}   \includegraphics[width=7.3cm]{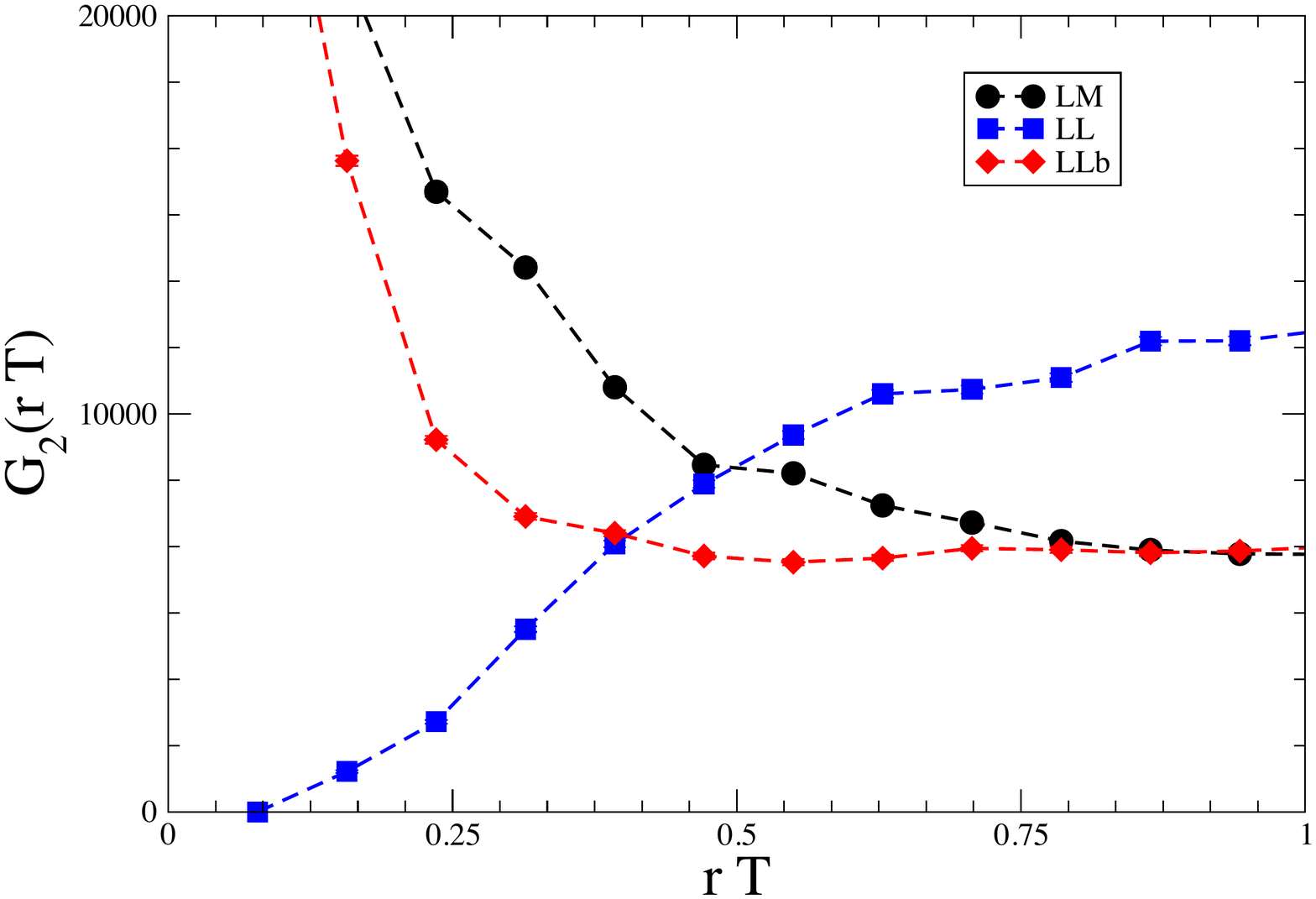}  
   \includegraphics[width=7.3cm]{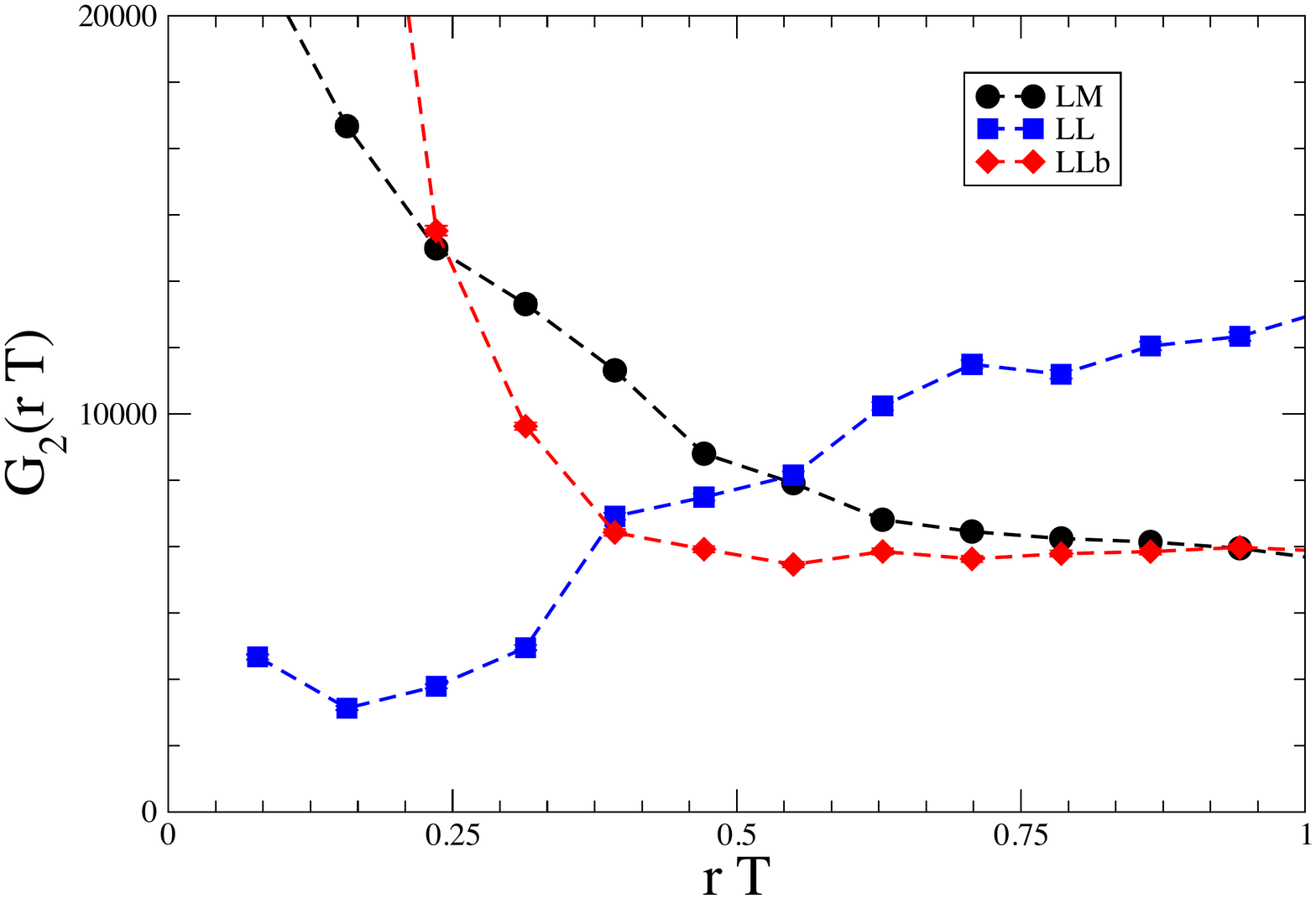}   \includegraphics[width=7.3cm]{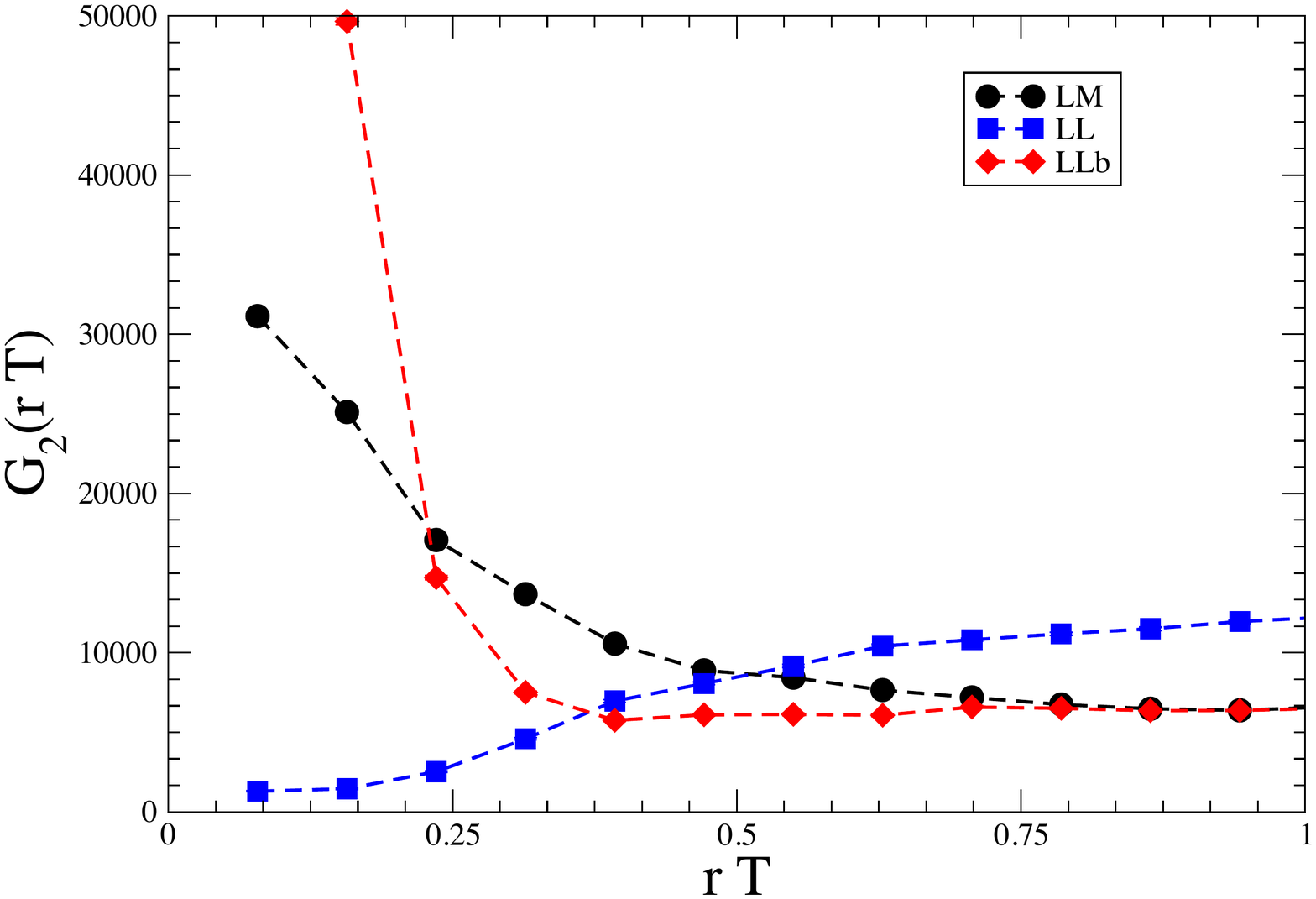} 
     \includegraphics[width=7.3cm]{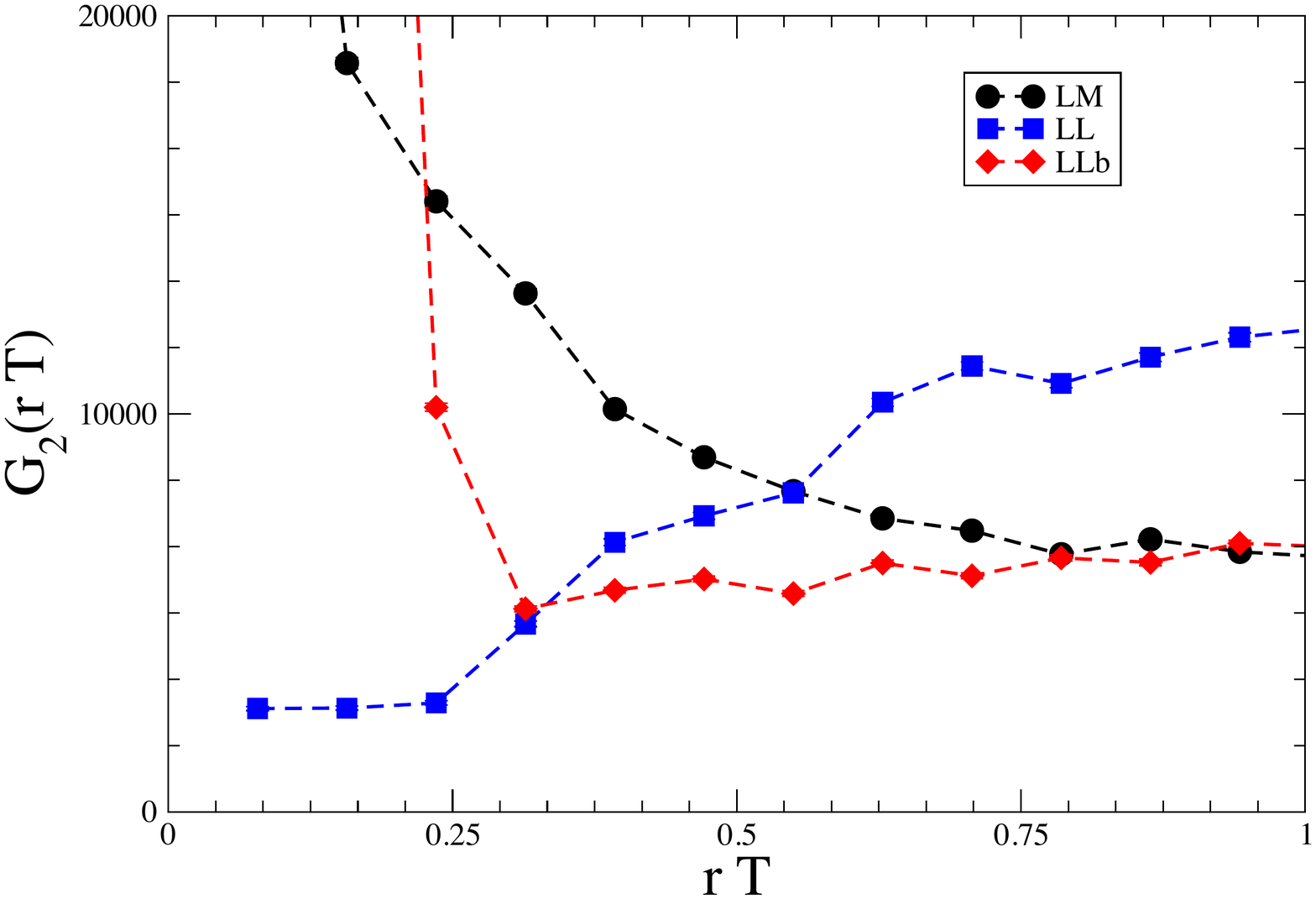}  \includegraphics[width=7.3cm]{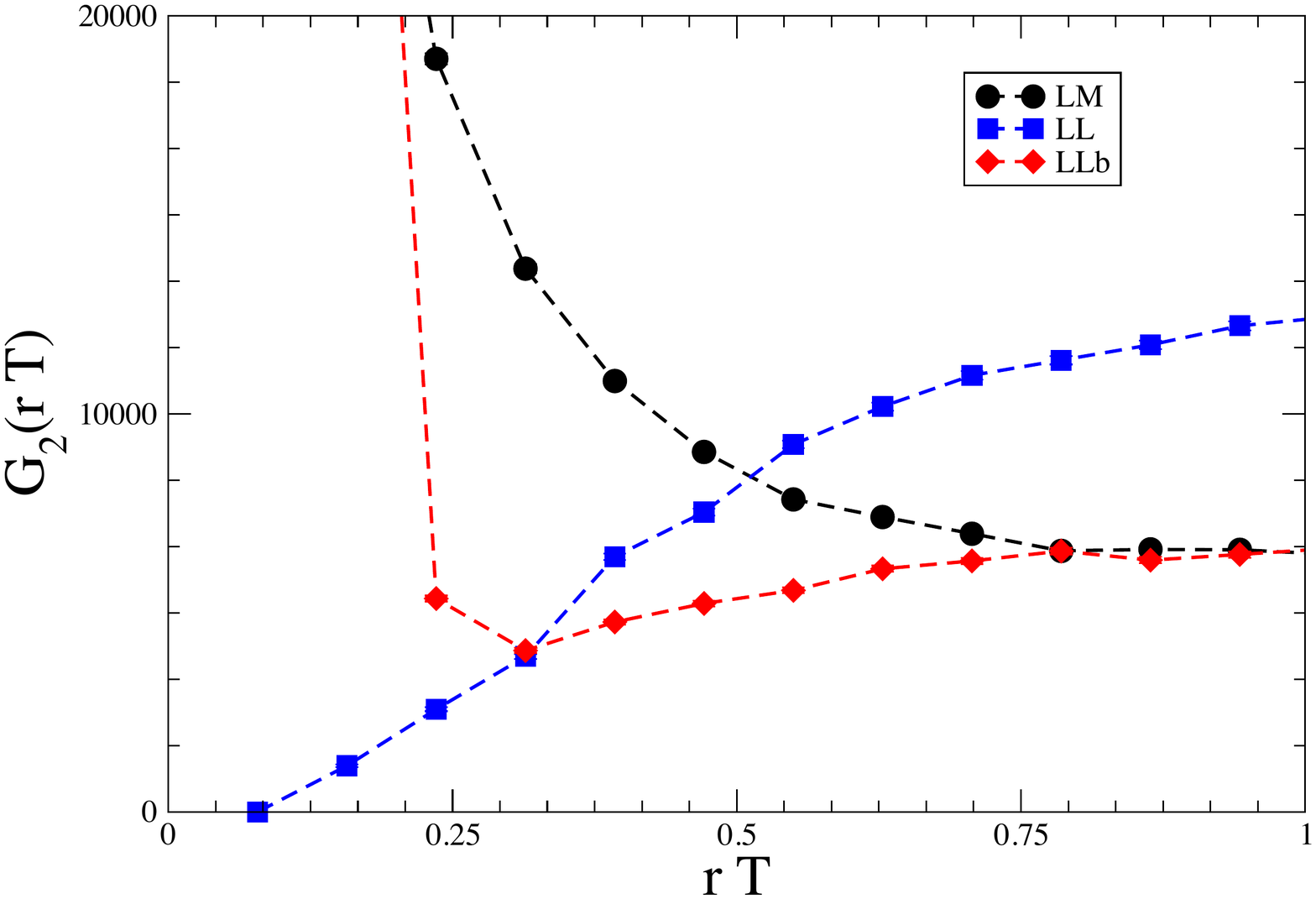}   
   \caption{   
  The correlation function for $LM, LL$ and $L\bar{L}$ dyons versus distance, normalized to the
  volume available. From top to bottom we show $N_f=1,2,4$, respectively.
  Left/right columns are for the  volumes per dyon  $VT^3=0.31, 1.04$.
     }
   \label{fig_corr}
\end{figure*}

Finally, about our units. Standard nonperturbative normalization of the gauge field eliminate the coupling
from the Yang-Mills equations. It appears in classical action $S=O(1/g^2)$, but since its total value depends on
the total number of dyons which is not changed in our configurations, it is absorbed in the overall input density.
The one-loop bosonic and fermionic determinants we actually simulate do not contain the coupling.  
Thus, there is no running and scale dependence in the model at all, 
and all quantities are essentially dimensionless.
Or, as we do, all quantities can 
   always be made dimensionless by multiplying by appropriate powers of temperature $T$.
   
In particular, the simulations are done for fixed number of the dyons on a  $S^3$ sphere,
with its radius characterized  by  the (dimensionless) radius  $R(S^3)T$,  or by the  proper volume per dyon in T-units, $VT^3$,
the inverse of the  diluteness (we remind that the total volume is $V(S^3)=2\pi^2R(S^3)^3$. 
 The values for which   the simulations has been done are listed in the
Table \ref{tab_runs}.
Note that the density or volume per dyon changes by about two orders of magnitude. The largest values
of the proper volume (per dyon) 
 correspond in physical ensemble to weaker coupling, larger dyon masses and  higher temperatures.
 Going down in this volume one goes into the denser systems, stronger coupling and
the lower $T$. Respectively we will monitor in simulations 
how all  observables depend on the proper volume,
focusing on the correlations between the dyons and the dirac eigenvalues
spectrum. 

\begin{table}[h!]
\begin{tabular}{|c| c| c|} \hline
 dyons &  $R(S^3)T$ & $ VT^3/dyon $ \\ \hline
 64 & 4.5 & 28.\\
 64 & 3.0 & 8.3\\
 64 & 2.5 & 4.8\\
  64 & 2.2 &  3.28\\
   64 & 1.5  & 1.04 \\ 
   64 & 1.2  &  0.53\\  
      64 & 1.  &  0.31  \\  \hline
\end{tabular} 
\caption{ The list  of the radii and volumes/dyon used in the simulations} \label{tab_runs}
\end{table}

 The only dimensional parameter which enters the fermionic
 determinant is the current quark mass $m$, which  in our units corresponds to the ratio $m/T$.
 Keeping its constant means a temperature-dependent mass. This does not represent a problem as long as one is interested in the massless limit, $m\rightarrow 0$, in which the classical action becomes scale invariant. Yet  in reality, following a prescription which is commonly adopted by 
 lattice practitioners, the actual
  value of this mass used in our simulations is not zero and is not 
 directly related to physical light quark masses: its role is to prevent influence of the finite volume effects. In particular, we shall will see below that,
in order to correctly monitor the chiral symmetry breaking, we are bound to consider masses  $m/T \gtrsim 0.2, $

\section{Simulation results} \label{sec_results}
One standard observable in the simulation is the acceptance probability
of Metropolis steps, which should be tuned to being comparable to rejection probability
to run it efficiently. 
 We estimate the number of Metropolis steps needed to achieve thermalization by standard auto-correlation analysis based on the dyons' position. Our typical runs include a total of 256 independent configurations, obtained from
 32 independent Markov chains, each consisting of  64000 Metropolis steps. All simulations were performed at the Interdisciplinary Laboratory for Computational Science (LISC), at Trento.

\subsection{The spatial correlations between dyons}
One of the benefits of going from 4-d torus to the $S^1\times S^3$ geometry is that in the latter case it is much easier to define the  interparticle distance 
 \be r_{i j} = R~ \alpha_{i j} =R~ \text{arcos}~ (\vec{n}_i \vec{n}_j ) \ee
where $cos(\delta_{ij})$ is the angle  between two points defined via the scalar product of their unit position vectors on $S^3$ and $R$ is its radius.
  
We collected histograms of the various two-particle correlation functions as a function of the (dimensionless) distance $RT$ on the sphere are normalized 
by dividing out the volume element $dV/d\delta_{ij}\sim sin^2(\delta_{ij})$ corresponding to random occupation on the $S^3$ sphere.
A sample of the results for different values of $N_f$ and different dyonic densities is shown in Fig.~\ref{fig_corr}. 
One can immediately identify a number of feature which are common to all such correlation functions:
\begin{itemize}
\item repulsive correlation between identical dyons,\\
\item attractive interaction between
the dyon and antidyon;\\
\item the radius of the correlation decreases as $N_f$ grows
\end{itemize}

 \begin{figure}[t!] 
   \includegraphics[width=7.3cm]{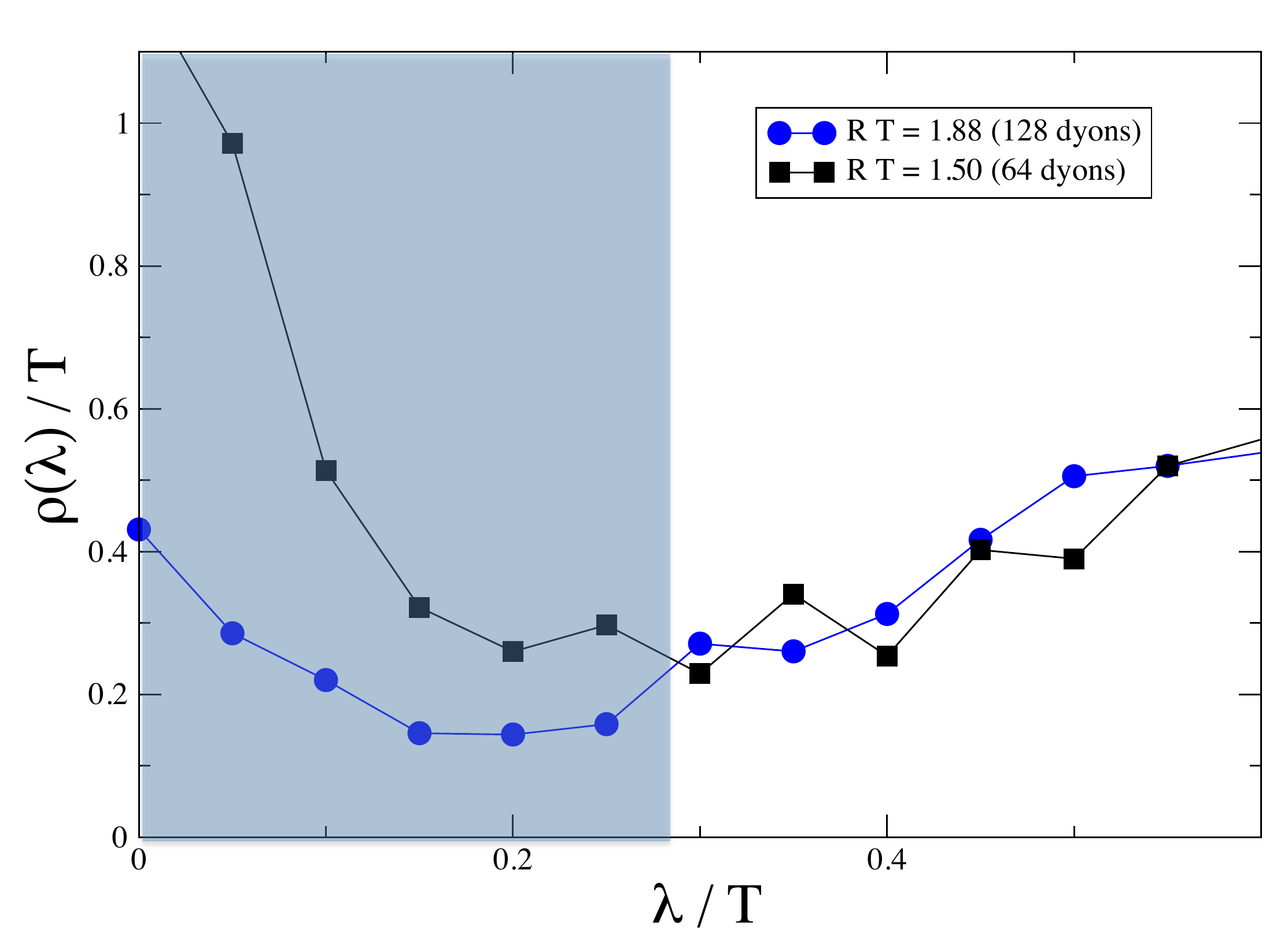}   
   \caption{   The spectrum of the Dirac eigenvalues in the quenched ensemble,  comparing
  two ensembles of 64 and 128 dyons at the same density, with zero current mass. The shaded area denotes the region of the spectrum
   where volume artifacts become significant. 
   }
   \label{fig_volume_mass}
\end{figure}

\subsection{The Dirac eigenvalues and the final size effects}

Our present computer resources allowed us to perform simulations with 64 dyons. 
In order to  estimate the magnitude of finite-volume effects, we performed also a few quenched runs with ensembles  consisting of 128 dyons.
   Two distributions of the Dirac eigenvalues 
  for 64 and 128 dyons performed at the same density conditions are compared in Fig.\ref{fig_volume_mass}.
  One can see that the finite volume effects are significant only for the smallest eigenvalues. More specifically,
  one finds distortions of the spectrum for the first
   bins, say for the shaded area 
   \be {|\lambda| \over T}< 0.25 \label{finite_V}\ee
 Notice also that that larger systems have smaller finite size effects, as expected.  
   
A rather crude account of the finite-volume effects can be thus made by simply disregarding the data points in
the strip identified by the inequality (\ref{finite_V}), yet we will need more refined fits when we introduce a nonzero quark mass in the next section. 

 \begin{figure*}[t] 
     \includegraphics[width=7.cm]{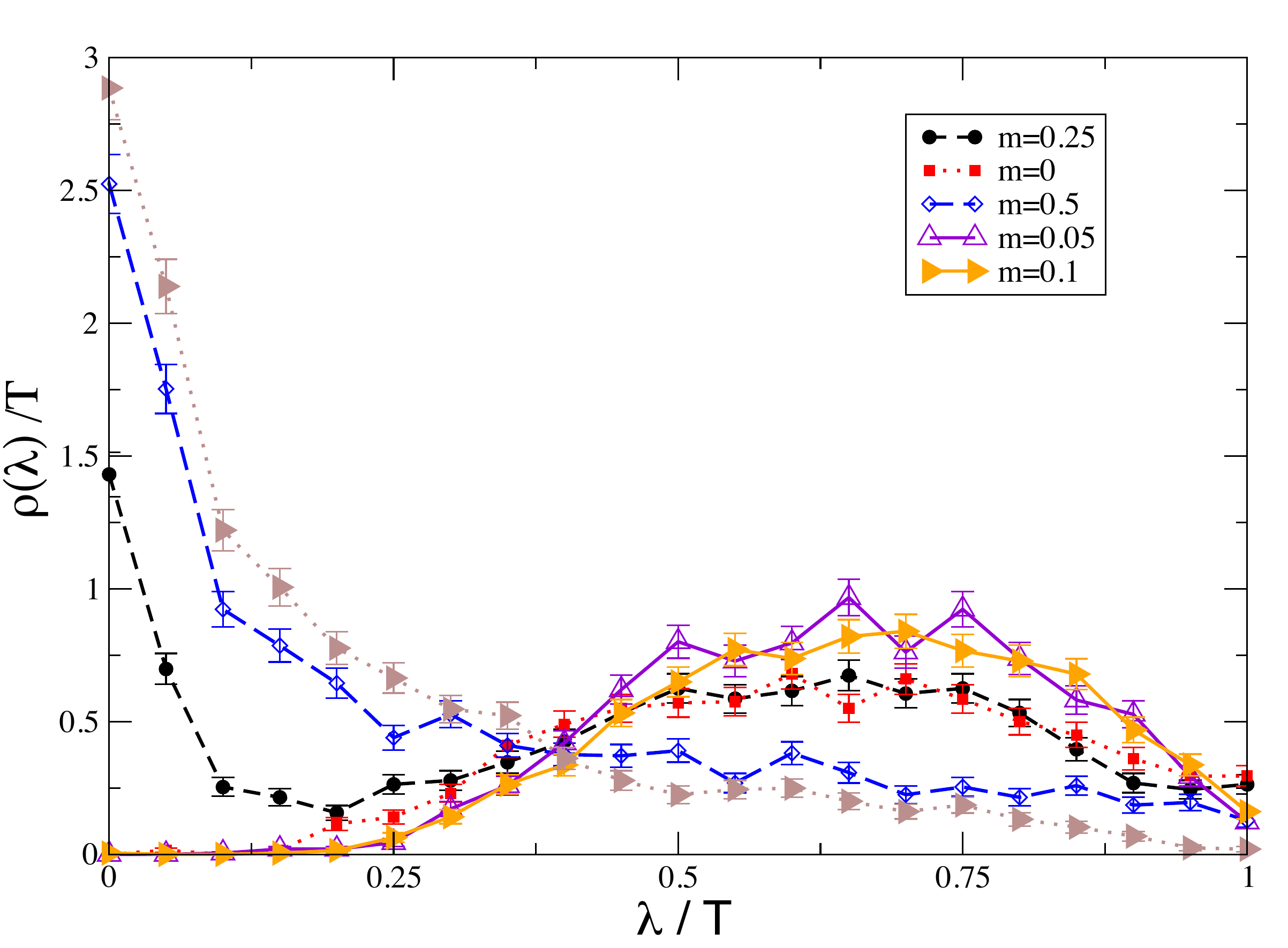}   
               \includegraphics[width=7.cm]{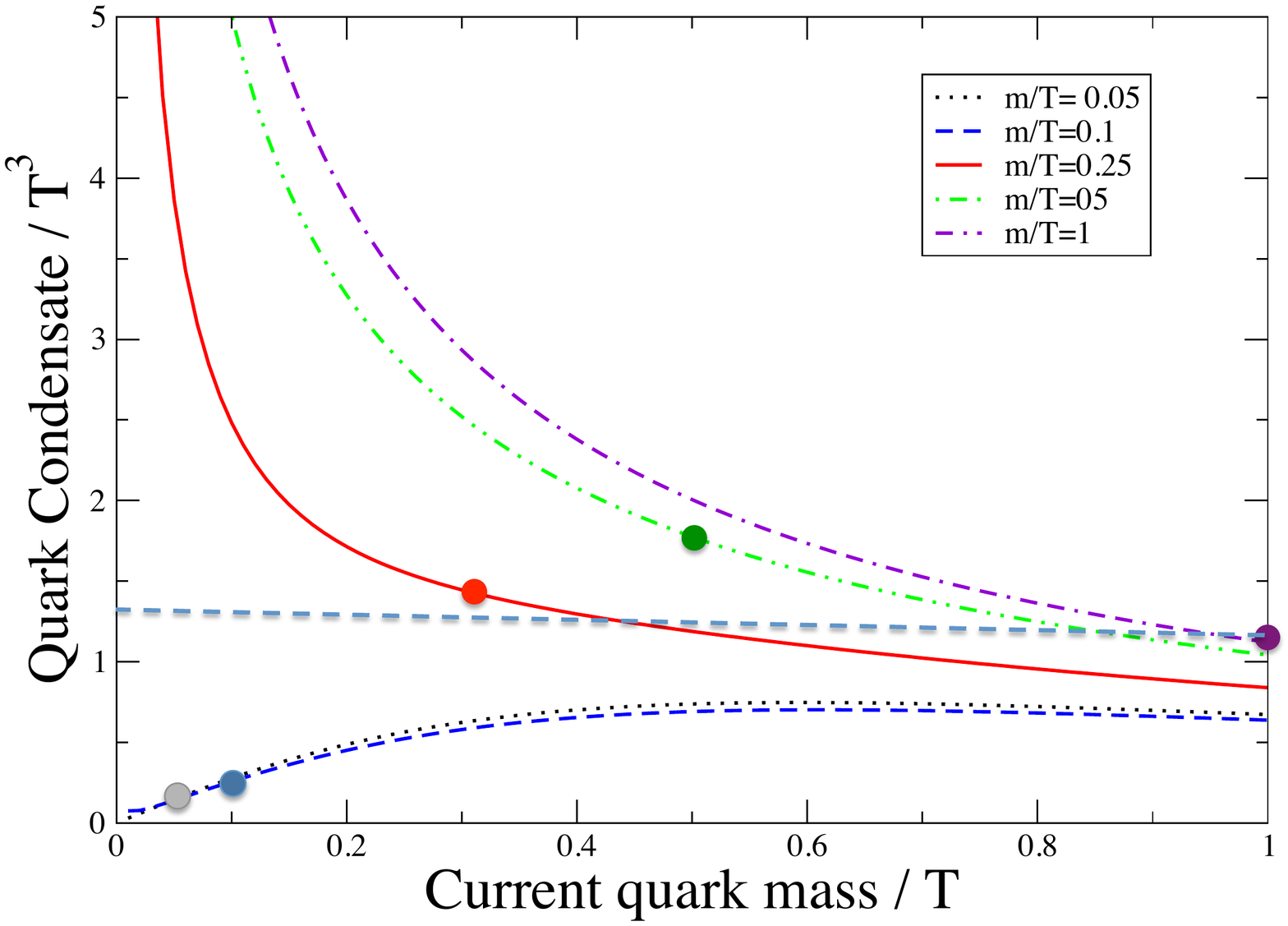}   
   \caption{  Left panel: Dirac spectrum as a function of the current quark mass $m$. Right panel: quark condensate as a function of the current quark mass obtained from the loop-estimate (\ref{eqn_loop}). The circles denotes the value of the mass used in the generation of the dyonic configurations. The simulations used for the results shown in both panels were performed on a $S_3$ of radius $1.5 T$ and with $N_f=2$.   }
 \label{fig_RL_mass}
\end{figure*}

\subsection{Quark condensate  dependence on the current quark mass}

    Let us start reminding that
   there are two types of quark masses: ``current mass" generated by 
   the interaction with the Standard Model scalar (Higgs) VEV with the left-right structure
   \be 
   L_{LR}=m ~(\bar{q}_L q_R+ \bar{q}_R q_L) 
   \ee
   and thus violates the chiral symmetry explicitly.    
   The other ``mass"  $M$ is generated by the holonomy. As it is due to the VEV
   of the vector field component $A_4$, its structure is chirally  diagonal (symmetric)
    \be L_{A_4}=M (\bar{q}_L q_L+ \bar{q}_R q_R) \ee
     We recall that $M$ appears in the parametrization of matrix elements of the $T$-matrix, which defines the Dirac operator in the background holonomy potential. (For clarity, we do not use
   the notation of ``constituent quark mass" induced by $sponateous$ chiral symmetry breaking
   in this paper.)

      \begin{figure*}[t!] 
        \includegraphics[width=6.3cm]{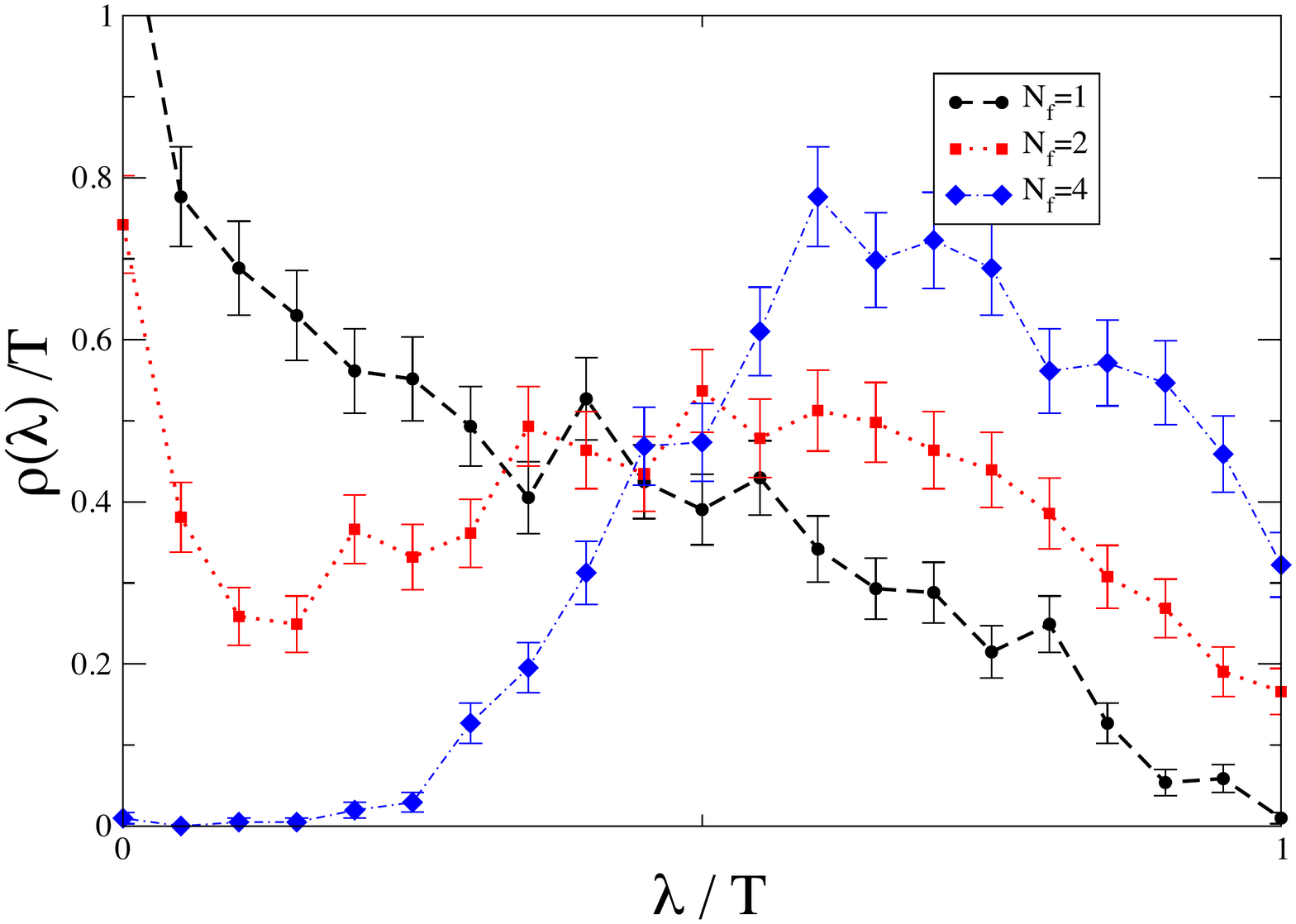}  
        \includegraphics[width=6.3cm]{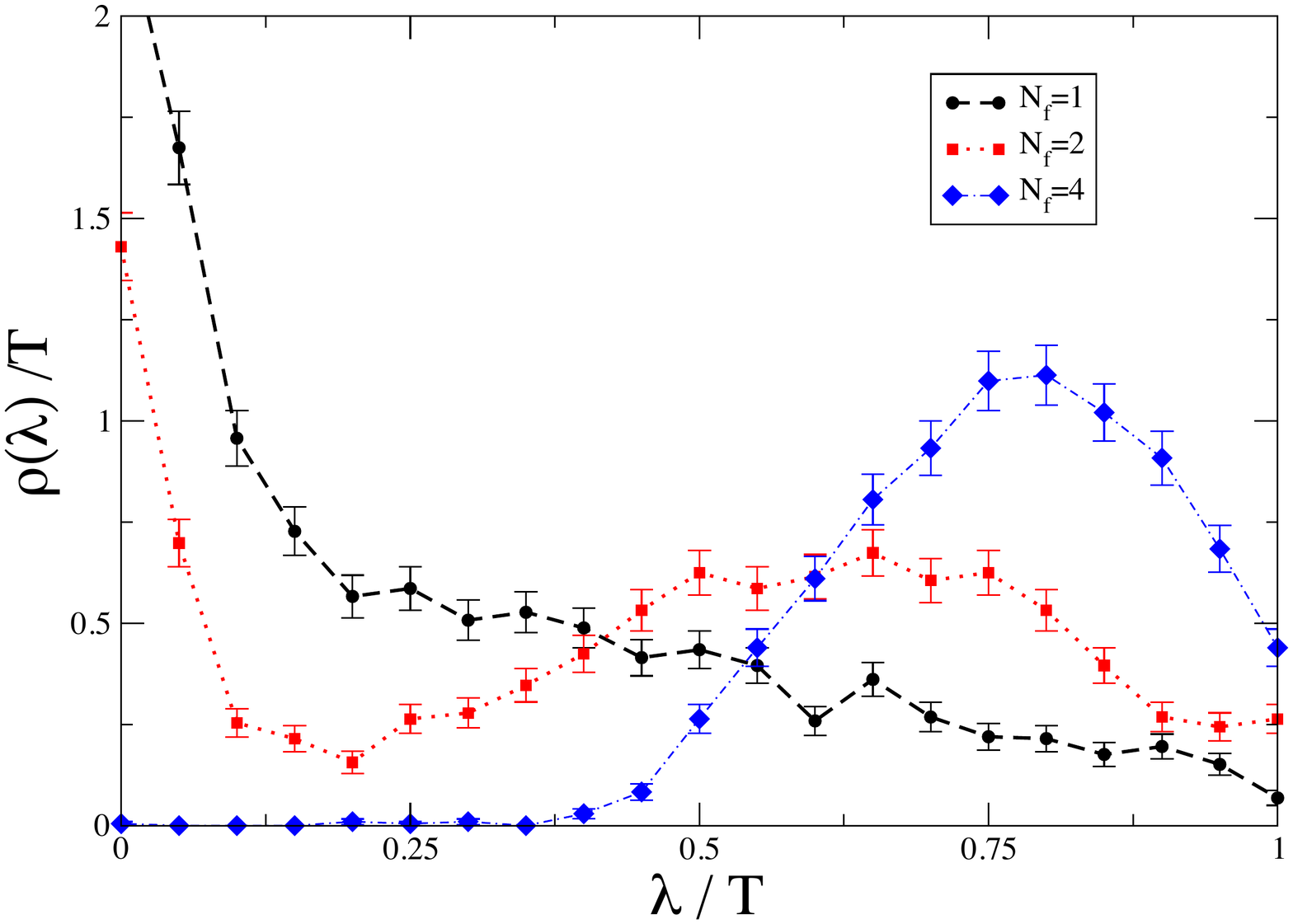}\\ 
                        \includegraphics[width=6.3cm]{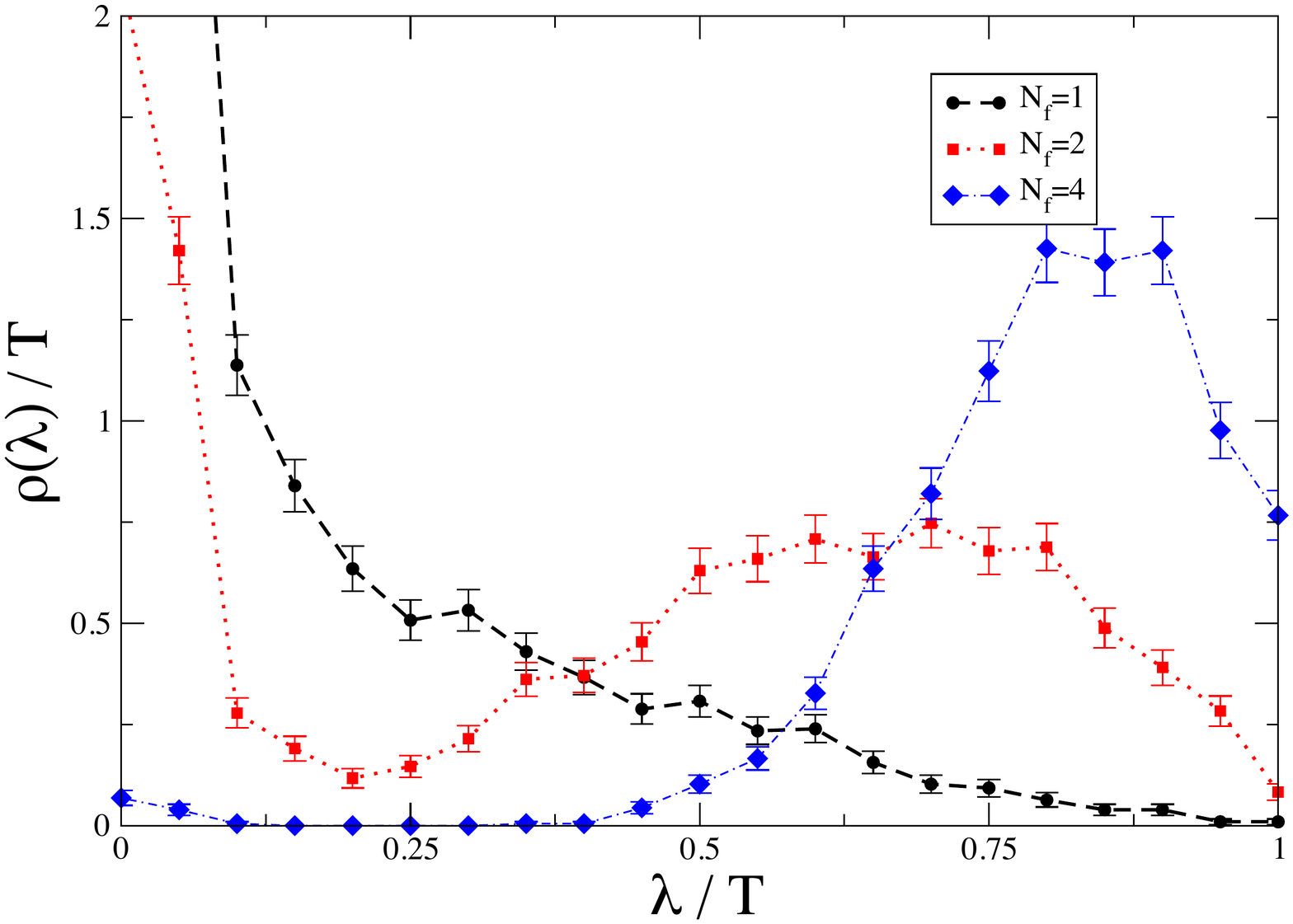}  
                      \includegraphics[width=6.3cm]{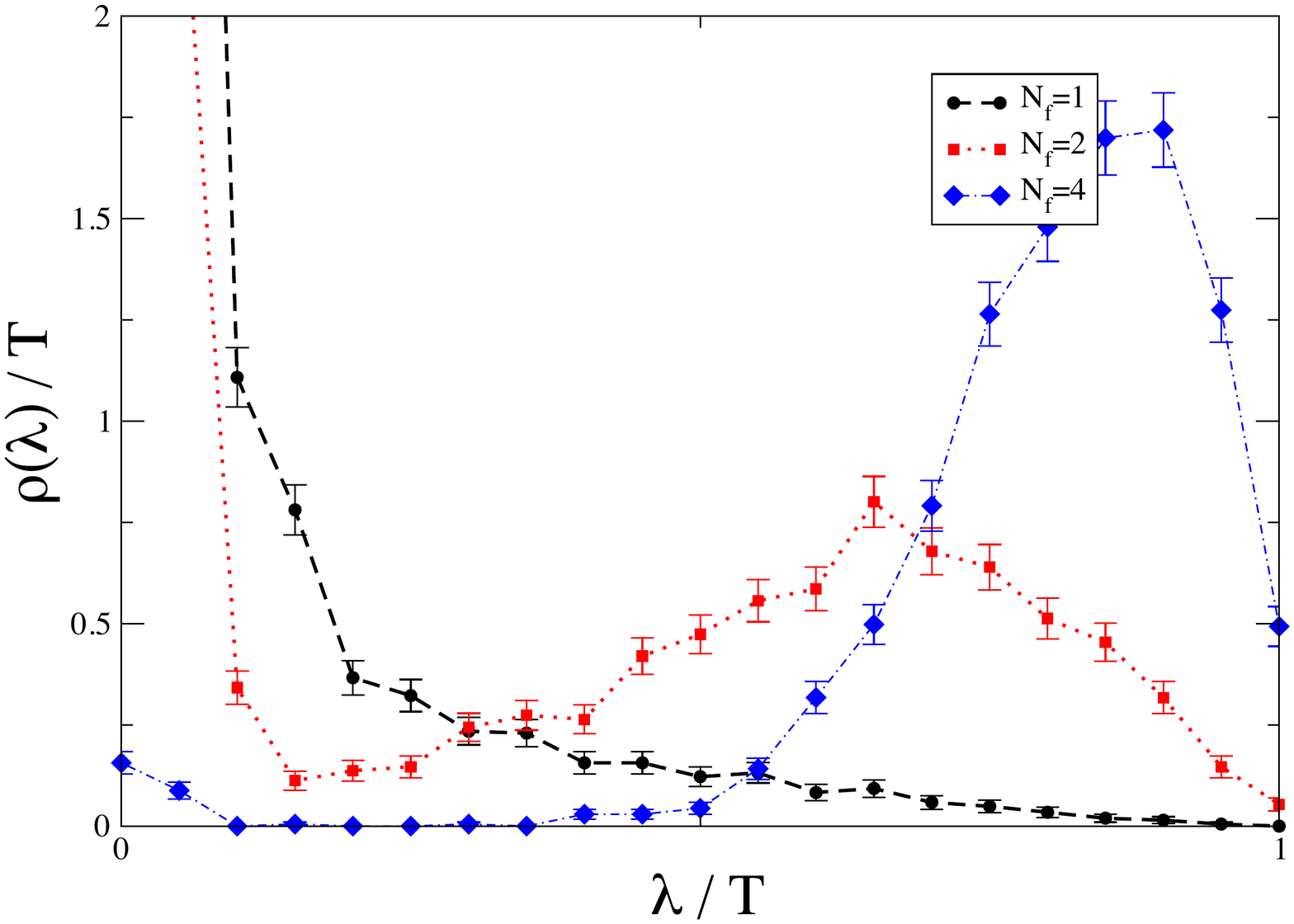}\\   
                                \includegraphics[width=6.3cm]{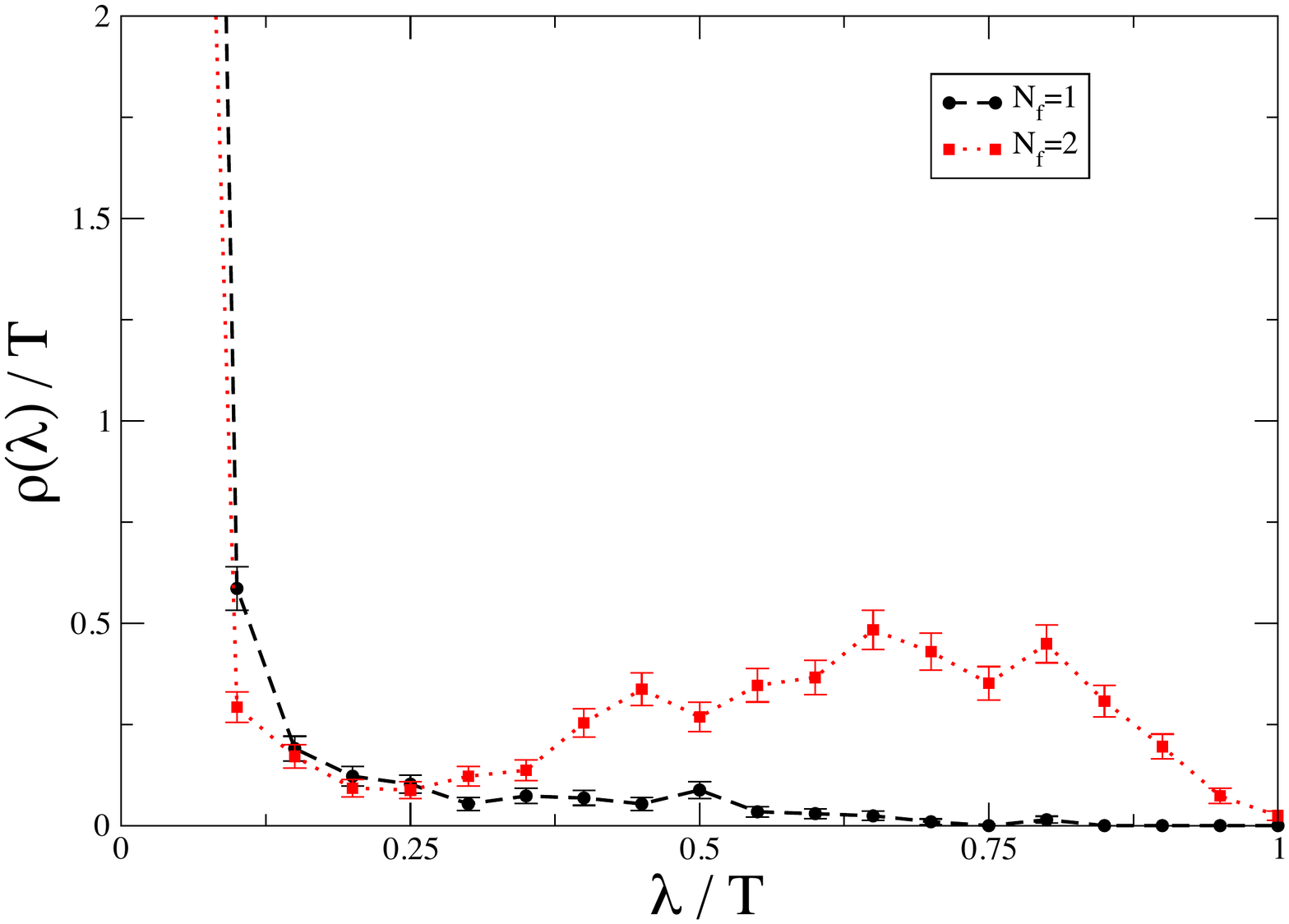}  
 \caption{ The Dirac eigenvalue distributions for three values of the $N_f=1,2,4$. The plots, ordered from top-left to right-bottom
  are performed on different radius of the 3-sphere (R=1.0~T,R=1.5~T ,R=2.2~T, R=3.0~T ,R=4.5~T), from high to low dyon density. The current quark mass was set to $m/T=0.25$.  }
   \label{fig_rho_with_mass}
\end{figure*}
   
   The current quark mass $m$  is  important for our study because    
    it  ``seeds"  the chiral symmetry breaking. 
   As it is well known to lattice practitioners,  in order to study  this spontaneously broken symmetry one formally
    should take the double limit, the infinite volume $V\rightarrow \infty$ and massless quarks $m \rightarrow 0$,
    in this order. In practice, all simulations are performed    
      on a finite volume, and the  absolute values of the smallest Dirac eigenvalues are bounded from below at the scale $\lambda \sim 1/V$. The
      current mass  should be selected somewhat above this scale, in order to cause the chiral symmetry breaking. The reason is obvious from 
      the  expression for the quark condensate, as the trace of the massive quark propagator
\be  
\la\bar{q}q\ra=  \sum_i {m\over \lambda_i^2+m^2} = m~ \int d\lambda ~{\rho(\lambda) \over \lambda^2+m^2}
\label{eqn_loop} \ee
   Tuning the quark mass values used in the simulation is important: it   should be as small as possible,
    yet large enough to avoid large finite-volume effects.

       An example of how this tuning  works in practice
        is demonstrated in Fig.\ref{fig_RL_mass}. In the left panel we show the Dirac eigenvalue distributions  
   for 6 values of the current quark mass. As one can see, the spectrum itself changes, from a dip near $\lambda=0$
   at zero mass, to a sharp peak at larger masses (as for quenched $N_f=0$ case).
    The  Casher-Banks ``density of the eigenvalues at the origin" thus changes dramatically, from zero to very large values:
    hence, reading off $\rho(0)$  is clearly not a good method for estimating the quark condensate estimate.  
   By contrast, the points shown in the right panel of  Fig.\ref{fig_RL_mass} show the quark condensate estimated from the loop expression (\ref{eqn_loop}). We observe that the quark condensate remains approximately mass independent for larger masses $m > 0.2 T$, while decreasing for
      smaller masses, eventually going to zero at $m=0$.
      
Let us first comment on the continuous lines  in Fig.\ref{fig_RL_mass}(b), passing through our 6 simulation points. Those  correspond
      to the loop expression (\ref{eqn_loop})  in which one uses the eigenvalues as determined from the simulation 
      with the corresponding dynamical mass, but in the loop formula  itself the mass is substituted by a  
 separate parameter, called the
``valence quark mass" $m_v$ in literature. 
Such ``valence approximation" has been used before, sometimes for quenched ensembles,
to approximate the mass dependence outside of the region where simulations were done.
 However, as seen from the plot, the trend indicated by those curves is quite misleading,
being completely different from what is obtained from the dynamical simulations -- the points  themselves. It shows that the main
mass dependence is stored in the eigenvalue distribution, reflecting the dynamical quark mass of the vacuum, of the ``sea" quarks.

Let us now remind the Smilga-Stern  theorem \cite{Smilga:1993in}, which is based on chiral perturbation theory and predicts the slope of the eigenvalue spectrum  near the origin, as a function of the chiral condensate $\la \bar q q\ra$, the pion decay constant $F_\pi$ and the number of flavors:
              \ba {dN\over d\lambda}(\lambda)=-{1\over \pi} \la\bar{q} q\ra\\ \nonumber + |\lambda | {(N_f^2-4) \over N_f} {\la\bar{q} q\ra ^2 \over 32\pi^2 F_\pi^4}+O(\lambda^2) \ea
Lattice and instanton liquid simulations had indeed confirmed that  the slope of the distribution changes sign at   $N_f=2$,
from a negative constant at  $N_f<2$ to positive at $N_f>2$. 

Let us also point out that the spectral density in the expanded form near $\lambda=0$ can be put into the integral in (\ref{eqn_loop}),
 generating the condensate expanded in powers of $m$. The zeroth order is the Casher-Banks term.
Returning to Fig.\ref{fig_RL_mass}(b) for $N_f=2$, we see that no liner term in $m$ is expected, and  higher order terms can be neglected: so the correct infinite volume $V\rightarrow \infty$ limit should be just a constant. Our guess of its value is indicated by a horizontal
dashed line.   Comparing it to the data we see that for $m/T<0.2$ finite solum effects are large and those should not be used.
Since we cannot afford to do simulations with variable masses in all cases studied, we will below only use 
the  dynamical mass $m/T=0.25$, at which we expect the condensate accuracy determination to be not worse than say 20\%. 

Our computed eigenvalue spectra, for different volumes and number of flavors,
are shown in Fig.\ref{fig_rho_with_mass}. While some finite-volume effect is still seen at the smallest eigenvalues, it is much less prominent than
in zero mass simulations discussed above: they are only seen for $\lambda/T\lesssim0.1$.

The $N_f=4$ spectra are clearly ``gapped", without any small eigenvalues, and the gap magnitude grows 
as the dyon density falls. We thus conclude that $N_f=4$ case needs rather large dyon density to break the chiral symmetry,
perhaps close to the highest one we used.

\begin{figure}[h!] 
       \includegraphics[width=7 cm]{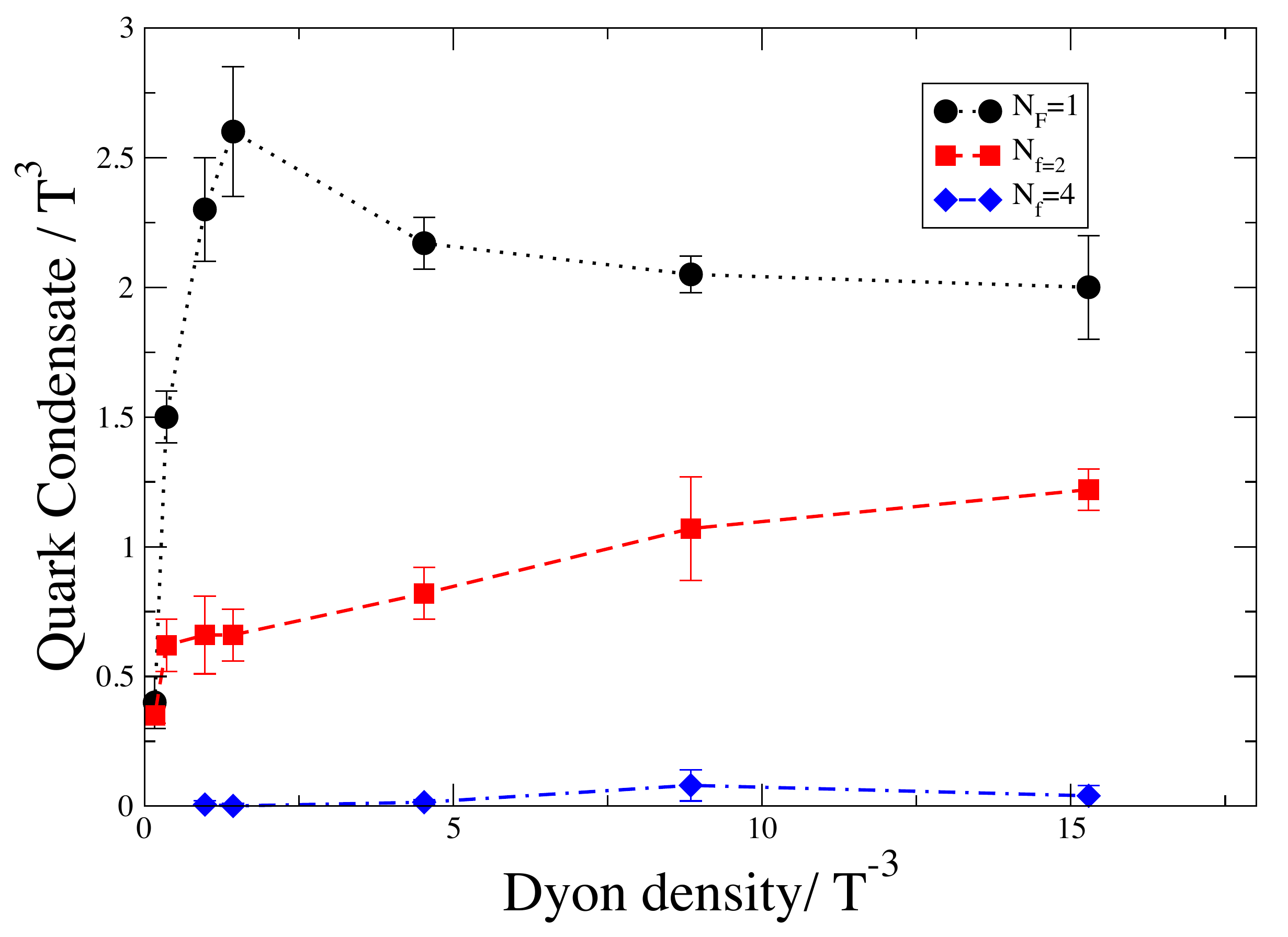}
   \caption{ The quark condensate versus the dyon density. The current quark 
   mass is set to $m=0.25~T$.  }
   \label{fig_cond_dens}
\end{figure}

Other $N_f$ simulation show no gap and thus correspond to chirally broken phase, except for very low dyon density.
At different dyon density and number of flavor, we performed  a linear fit $\rho(\lambda)=Q+C\lambda$.
The fit was restricted to a range with $0.1<\lambda/T$ until the maximum $\lambda$ for which the reduced $\chi^2$ remains smaller than 1. 
This way, we identified constants $Q$ and $C$ with the quark condensate and
the Smilga-Stern constant, respectively.
The result of the fit is shown in  Fig.\ref{fig_cond_dens}.
The nonzero condensate observed for 
 the $N_f=1$ and $N_f=2$ ensembles at high dyon density is rather density-independent.

  More systematic account for the finite-volume effects and more accurate
  determination of the condensate can be done by comparing our data with
   the expectations for the so called 
  ``mesoscopic regime", in which the volume is not macroscopically large.
Quantitative predictions of the shape are known from  
  chiral random matrix models \cite{Shuryak:1992pi}, which are well confirmed on
  the lattice. We plan to do so elsewhere.

\subsection{Free energy, dyons' interaction,  the back-reaction and confinement}

The well known perturbative holonomy potential, 
which has a minimum at zero holonomy, has been argued by Polyakov's original work to get
cancelled by the  nonperturbative effects, resulting in a vanishing Polyakov line and thus confinement.  
Lattice study of the effective holonomy potential, in particularly recent Ref.\cite{Diakonov:2012dx}, 
have indeed found such behavior. 
 Diakonov \cite{Diakonov:2009ln} further argued that the nonperturbative contribution comes from the back
reaction of the  dyons. He also argued that (at least the Coulomb-like moduli part of) the dyon interaction
is small and can be approximately ignored.
In this section we make the first step toward understanding of whether those ideas are correct.

\begin{figure}[t] 
   \centering
   \includegraphics[width=7cm]{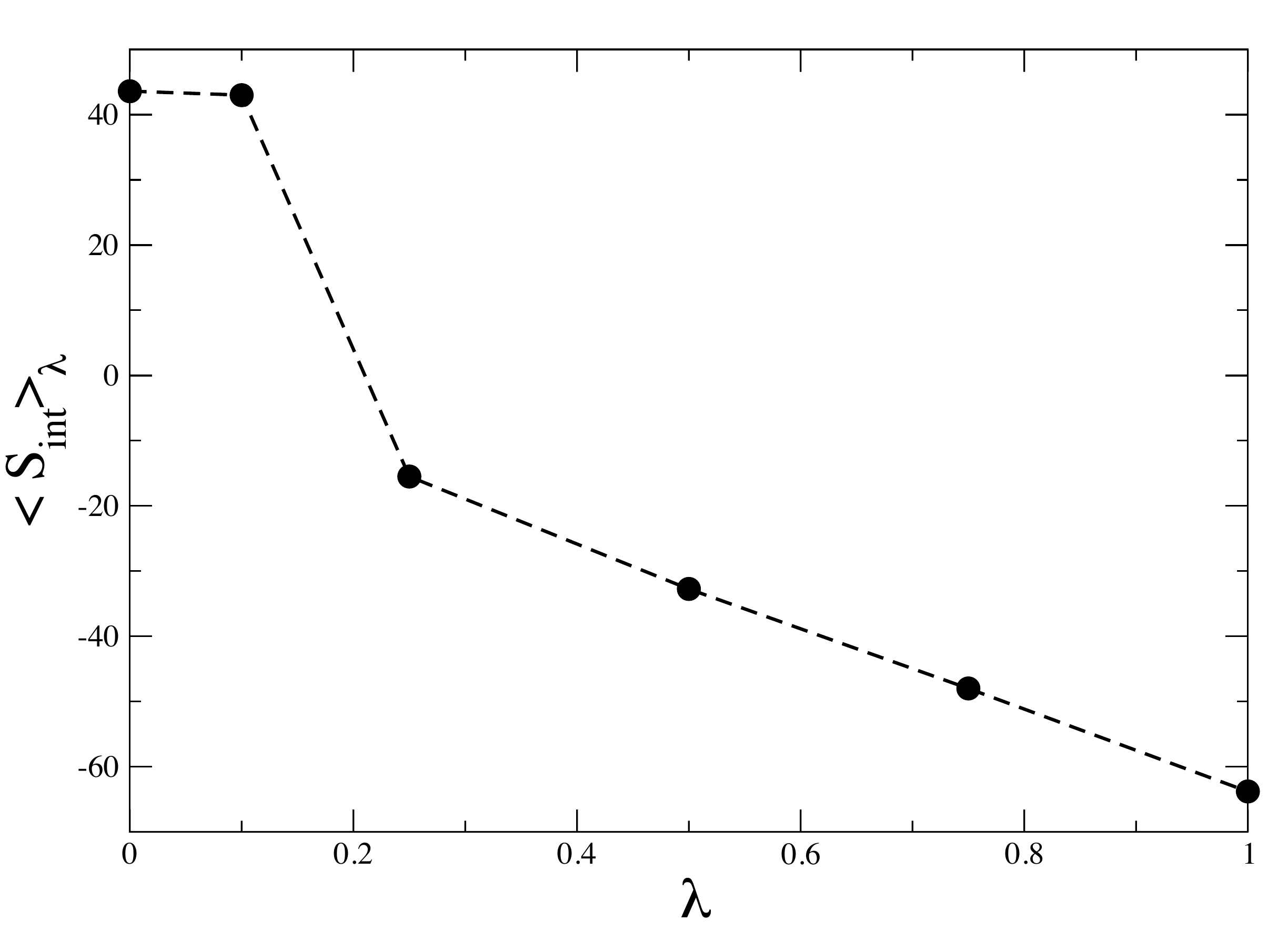} 
   \caption{The dependence of the average interaction free energy on the dimensionless adiabatic parameter $\lambda$.
   \label{fig:free_energy_dyons}
   }
\end{figure}

We calculate the change in the free energy of the system, between the interacting and  noninteracting dyon gas using standard
a thermodynamic integration method, based on the adiabatic switching of the interaction between the particles. 
One splits the action into independent particles and
their interaction, introducing  the adiabatic parameter $\lambda$
\be S=S_0+\lambda S_{int} \ee
and do simulations with $\lambda$ changing from 0 to 1. The resulting free-energy is recovered as follows
\be F=F_0+\int_0^1 d\lambda < S_{int}>|_\lambda \ee
The resulting dependence of the action as a function of the adiabatic parameter is shown in Fig.\ref{fig:free_energy_dyons}.
Of course, we include all three ingredients of the interaction, the moduli, screening and the fermions,
and measure them separately as well: but for brevity we will not discuss those details here.

As one simulates an ensemble with a fixed number of dyons, one can disregard any constant factors in the partition function,
as the relative weights of the ensemble configurations depends only on terms which
are functions of the dyonic collective coordinates. 
This explains why one has positive  free-energy of the noninteracting ensemble at $\lambda=0$, which is  not really physical
and depends of what factors we do or do not include in the free gas expression.
The free-energy change $\Delta F=-22$ is physical, and a shift to negative is typical for liquids.
 The value itself is for the entire system of 64 dyons: it thus corresponds to a free energy change per particle of 
$\Delta F/N=-0.34$ which is due to the attraction in the system.

To put it in perspective, one should know the action per dyon. We will discuss available lattice data in the Appendix:
the effective  action fitted to those are about $S\approx 3$ per dyon. The result of this section show that the average interaction
part of it is reasonably small, $\sim 1/10$. On the other hand, the partition function is enhanced by about
$exp(.34)=1.4$ per dyon, not a negligible enhancement.

As for the back reaction to the total holonomy potential, we think that it is premature to
discuss it now, before lattice simulations provide more define estimates on the dyon density.

\section{Summary and further developments} \label{sec_summary}

   This paper is the first study of the topology, in the finite-temperature QCD near $T_c$, which incorporates the nonzero
   holonomy and the (anti)self dual dyons as the ingredients. We (i) formulate
the partition function for (anti) self-dual dyons incorporating one-loop screening, moduli space measure and sermonic
zero modes; 
and (ii) perform the first direct      
   simulation of the resulting statistical ensembles.
We obtain certain set of results, showing how the dyon system depends on  the dyon density (changed by about 2 orders of magnitude), 
 on the number of light quark flavors
 $N_f$, from 0 to 4, and on the (current) quark mass $m$. 
 We have in particularly studied the Dirac eigenvalue spectra and  chiral properties: the gaps 
 in the symmetric phase and the quark condensate in the broken phase. More generally, we may say that
 in spite of the long-range nature of Coulomb forces involved and linearly rising screening,
 the simulations of the dyonic system turned out to be not as challenging as we thought when we started this work.
While the correlations between constituents are locally strong, overall liquid-like behavior is observed.

At the moment we are not calculating the fermionic propagators and correlation functions of operators with various
quantum numbers, but it can be done rather directly, in a way
similar to that in the instanton liquid. 

Mapping those results to finite-T QCD can be done provided the information about the dyons,
such as their density we reviews briefly in the Appendix, are extracted from the lattice. 
One should also directly relate fermionic quasi zero modes with the underlying topology,
which we believe will display the dyonic dominance in the chiral symmetry breaking phenomenon,
in the near-$T_c$ region. One may also expect progress in respect with the back reaction of the dyons on the holonomy
potential and the origin of confinement.

Another  direction is to extend the model toward  theories in which one has
high-T confinement under the analytical control, see e.g. \cite{Poppitz:2012nz}, by  including adjoint fermions  
with modified (e.g. periodic or with an arbitrary phase) boundary conditions on the torus.

Extending our results to  lower temperatures -- or much smaller holonomy values -- remains a numerically challenging task. 
Attractive correlations of the  LM  pairs should mostly recover the original instantons, yet preserving
small  Polyakov line value and confinement. 
At the same time, fermions induce the $L\bar{L}$ attractive interaction, producing topologically neutral clusters
whose role in the QCD vacuum has not been elucidated or studied.  \\ \\




{\bf Acknowledgements} The first  acknowledgements on this subject  obviously goes to Pierre van Baal and his students, who pioneered
the instanton-to-dyon transition at nonzero holonomy. Another source of inspiration came 
from  beautiful works of  Dmitry Diakonov, who sadly is no longer with us.
 ES acknowledges the role of his collaborator
and former student  Tin Sulejmanpasic. The work of ES is supported by  the US-DOE grant DE-FG-88ER40388. The work of PF was partially performed at the Interdisciplinary Laboratory for Computational Science (LISC), a joint venture of Trento University and Bruno Kessler Fundation. 

\appendix

\section{Distances on the $S_3$ sphere}
In our program, all distances are defined on a 3-dimensional sphere embedded in 4 dimensions.  Let $z_{n}^i$ be a unitary vector in 4 dimension, which will be associated to the position 
of the pseudo-particle of type $i$ in the $n-$th molecule. 
Monte Carlo trial moves are made by performing random rotations of given small angle (boldness) on the sphere. 

The angle between the 4-D vector $z_n^i $ identifying the pseudo-particle of type $i$ relative to the molecule $n$ and $z_m^j$ that points to 
 the pseudo-particle of type $j$ relative to the molecule $m$ on the unitary sphere  is given by:
\ba
\alpha_{i n,j m} &=& \arccos\left[ z_n^i \cdot z_m^j\right]
 \ea
(Notice that this formula is ill-defined for $i=j\cap n,m$. On the other hand, the distance of a particle from itself is never needed in the calculation.)
Hence the distance between these pseudo particle is simply:
\be
r_{ i n,j m} = a~\alpha_{i n,j m},
\ee
where $a$ is a length scale which can be identified with the box size.

For illustration purposes we mapped the $S^3$ to a flat $R^3$ space
using the stereographic projection. 
The three dimensional coordinates of our dyons are defined on a sphere, hence we use 4-dim vectors with the constraint
$n_1^2+n_2^2+n_3^2+n_4^2=R^2$. 

One may transform them into the 3 polar angle angles like this
\ba 
n_1 &=&R~\cos(\psi)\nonumber\\
n_2 &=&R~\sin(\psi) \cos(\theta) \nonumber \\
n_3 &=&R~\sin(\psi) \sin(\theta)\cos(\phi) \nonumber\\
n_4 &=&R~\sin(\psi) \sin(\theta)\sin(\phi)
\ea
The inverse formulas are:
\ba
\theta &=& \text{cotg}^{-1}\left(\frac{n_2}{\sqrt{n_4^2+n_3^2}}\right)\nonumber \\
\phi &=& 2 ~\text{cotg}^{-1}\left(\frac{\sqrt{n_4^2+n_3^2} + n_3}{n_4}\right)\nonumber \\
\psi &=& \text{cotg}^{-1}\left(\frac{n_1}{\sqrt{n_2^2+n_3^2+n_4^2}}\right)
\ea

In a stereographic projection to the flat 3d-space in polar coordinates $r,\theta,\phi$
the $\theta$ and $\phi$ angle keep the same 
meaning as above, while the distance from the origin $r$ is defined by the angle
$\psi$ according to
\be r= 2~R \tan(\pi/2-\psi/2) 
\ee
(R=1 is the sphere radius and the argument of tan is the angle conjugated to psi in the equilateral triangle of the projection).

Note that $\psi=\pi$ maps the south pole to the origin $r=0$, and the equator $\psi=\pi/2$
is mapped to $r=2R$. Hence the cartesian coordinates on the projected flat space are:
\ba
x_1 &=&  2~R~ \tan(\pi/2-\psi/2)~ \sin(\theta) \cos(\phi)\nonumber\\
x_2 &=&  2~R ~\tan(\pi/2-\psi/2)~ \sin(\theta) \sin(\phi)\nonumber\\
x_3 &=& x_1=  2~R ~\tan(\pi/2-\psi/2)~ \cos(\theta) 
\ea

\section{The Coulomb fields on a $S^3$} \label{app_Coulomb}

The usual coordinates $\psi,\theta,\phi$ 
have the following line element (metric) 
\be dl^2=d\psi^2+ sin(\psi)^2 d\theta^2 + sin(\psi)^2\sin(\theta)^2 d\phi^2  \ee
Standard Poisson equation is
\be {1 \over \sqrt{g}} \partial_\mu  \sqrt{g} g^{\mu\nu} \partial_\nu f=4\pi \rho \ee
For a point charge at the north pole ($\psi=0$)
the solution  is obviously independent on $\theta,\phi$ and we need to solve only
the $\psi$ -dependent part
\be f"(\psi)+2 f'(\psi)/\tan(\psi) =4\pi \rho
  \ee
  For zero r.h.s. one get solution 
  $f=(1/4\Pi \tan(\psi) $ which looks like a
  positive charge at one pole and the negative charge at the other. Indeed,
  on a compact manyfold all field lines have to go somewhere, and symmetry require it to
  be the opposite pole. 
  
  One remedy can be introducing the homogeneous compensating charge. 
  This solution is
  \be f={1 \over 4\pi}~\left[  1-{(\psi/\pi-1) \over \text{tan}(\psi)}\right]
  \ee
  It is compared with the naive Coulomb-like
  $1/\psi$ and solution given above in Fig.
  One can see from it that its behavior is very reasonable, and it vanishes at the opposite pole. 
  
  Screening effect contains the averaging of the dipole potential over the 3-d space,
  which produces the linear rising effective potential (as explained in the text). Keeping
  one charge at the north pole and spreading another at $\psi=\alpha$ 2d sphere one gets
  another solution of the Poisson eqn. It is 
  \be A(\psi<\alpha)= {1 \over 4~\tan(\psi)\pi}-{1 \over 4~\tan(\alpha) \pi} \ee 
  with $A(\psi>\alpha)=0$. Integrating the square of this field over the 3d sphere one gets the following expression
  \be 
  \int_0^\alpha d\psi A^2 \sin(\psi)^2= {\alpha -
  \sin(\alpha)~\cos(\alpha) \over 32 \pi^2~\sin(\alpha)^2} 
   \ee
   Both  agree reasonably well till about the equator of the 3d sphere.

\section{SU(2) holonomy on the lattice}
\label{Dyonlattice1}

 One important potential input  is the holonomy value $v(T)$ (i.e. the
VEV of the Polyakov line's VEV $\langle P\rangle $). While we have not yet used it, we indicate
that its  temperature dependence for the SU(2) gauge group is known. To our knowledge
the most  accurately determined
values are from Ref. ~\cite{Hubner:2008ef}.
The resulting
 fit of the Polyakov line VEV
  \be 
\la P(T)\ra = 1.22~t^{0.326}~(1-0.246~t^{0.53}), 
\ee
where $t$ is  the  near-critical parameter \be t=(T-T_c)/T_c  \ee
 The temperature dependence of the holonomy
is then $\textrm{arccos}\la P\ra=v(T)/2T$.

Note that these data and the fit are
consistent with the indices of the Ising universality class to which the SU(2) pure gauge theory
belongs.

\section{The (anti)self-dual dyons on the lattice}
\label{Dyonlattice2}

While we only consider a case with 2 colors,
we remind the reader that one of the major benefits of the dyon approach is the smooth behavior of the action in the large-$N_c$ limit.
Indeed, the dyon partition function is roughly the $1/N_c$ root of the instanton partition function, so
the dyon action in the large-$N_c$ limit is expressed via 't Hooft coupling $\lambda=g^2 N_c$ as
\be S= {8\pi^2 \over g^2 N_c}= {8\pi^2 \over \lambda} \ee
While at very strong coupling $\lambda\rightarrow \infty$ it is small, at the ``physical" value of the parameter $\lambda=20-30$
usually used for strongly coupled QGP applications it is still in the range of 3-4 $\hbar$-units.
Hence,  the weight factor $\exp(-S)$ is still a rather small number, providing a  substantial suppression of the dyon density.
It is important to stress that, although in principle it may be argued that the value of the action $S_0\approx 3-4$
is not large enough to justify the usage of the semiclassical approach,  
 the corresponding dyon ensemble  remains dilute even close to $T_c$, hence providing an \emph{a posteriori} justification of the approach.

  While we generally postpone mapping of the results of our dyonic model to
  the QCD thermodynamics directly,
in this section we still discuss some available results on the dyons, from
Refs.   \cite{Ilgenfritz:2006ju}
 and \cite{Bornyakov:2008im}. One of the issues here is whether they are or are not
   compatible with the with
  semiclassical expressions, in spite of possible low accuracy of those.
   
 These works both
 investigate the SU(2) gluodynamics on the $20^2\times 6$ and  $20^2\times 4$ lattices respectively, adopting the same tree-level improved Symanzik action at a bare coupling $ \beta_c=2N_c/g^2=3.248$.
In the former lattice, this choice corresponds to the critical temperature,  $T=T_c$, while in the latter lattice to a temperature  $T=(3/2)T_c$.

Note that lattice practitioners often introduce 
 \emph{ad-hoc} units of energy and length which are defined in such a way to mimic the dynamics of the physical world.  For example, 
In order to fix an energy scale, they measure the dimensionless constant.
\be 
\sqrt{\sigma(T=0)}a =0.236, \,\,\, T_c/\sqrt{\sigma(T=0)}=0.71
\ee
and artificially set the vacuum string tension  $\sigma(T=0)$ value to $(420 \, \text{MeV})^2$, i.e. to 
 its  physical value in the 'real-world' QCD. Furthermore, one keeps also
the 'real-world' relation between the length and the energy units
 $\text{fm}=1/(197\, \text{MeV})$.  Under this set of  conventions, one obtains the following ``absolute" values for the lattice spacing and
 the critical temperature
 \begin{eqnarray}
 a &=& 0.236 \times (197/420) \simeq 0.11 \text{fm} \\
 T_c &=& 0.71 \times 420\approx 300 \text{MeV}. 
 \end{eqnarray}

 When comparing to lattice results from those works -- such as the density of various dyons -- we may  
 adopt such absolute units. However, in general, we will refrain from doing so because this may lead to confusion when studying the dependence on the
  number of  flavors $N_f$. In addition, holding fixed the value of  the string tension constant is only practical
  in a lattice approach, in which this observable can be straightforwardly calculated.  In view of such considerations, in the present analysis we avoid using ``physical units" and consider only dimensionless ratios and couplings.
 
 \begin{figure*}[t] 
   \centering
   \includegraphics[width=7.3cm]{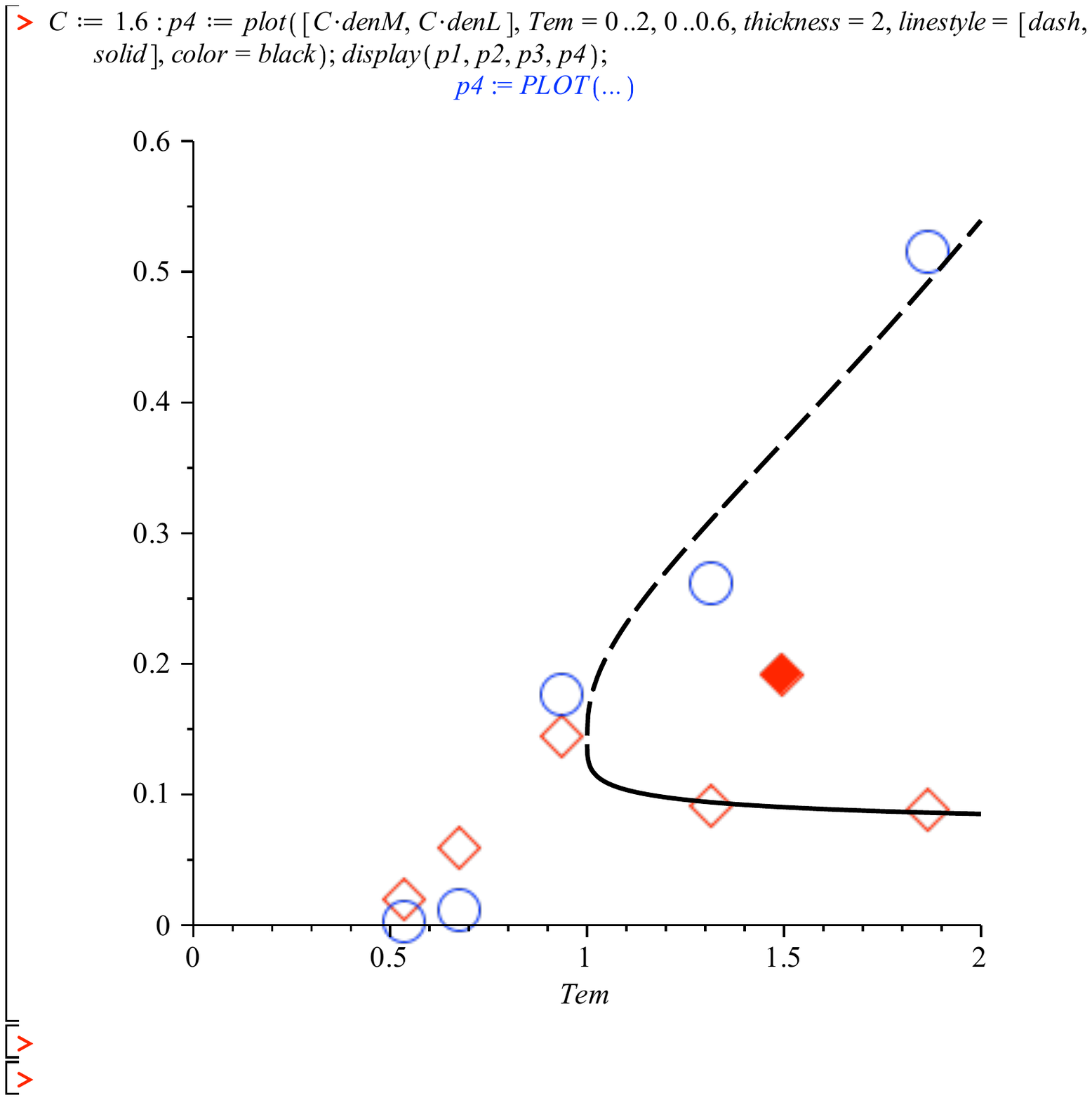}   
   \caption{   (color online)
  The density of the topological clusters (in $fm^{-4}$) versus the temperature $T/T_c$ of the SU(2)
  pure gauge theory. The open blue circles show static dyons, identified as $M$-type, while the open
  red diamonds are for calorons or $L$-type dyons \cite{Ilgenfritz:2006ju}. The closed diamond at
  $T/T_c=1.5$ is from \cite{Bornyakov:2008im}, in which $L$-type dyons were identified directly
  by fermionic zero-modes and the value of the Polyakov loop at its center. The dashed and solid lines
  correspond to semiclassical expectations for $M,L$ dyon density, with parameters defined in the text.
     }
   \label{fig_dens}
\end{figure*}

In Ref.\cite{Ilgenfritz:2006ju}, Ilgenfritz and co-workers report a list of the absolute densities
of static ($M$-type) dyons  and calorons ($LM$ dyon pairs) which were found in their lattice calculation.  
As $M$ dyons are much lighter than $L$, the calorons density is dominated by
 the density of the heaviest $L$-dyons.
 Their results are plotted in Fig.\ref{fig_dens}. In a more recent  paper by Bornyakov and co-workers  \cite{Bornyakov:2008im} an improved
identification of the dyon type was preformed by adopting the fermion method and also by computing the value of the Polyakov loop at the dyon center: it should be 1 for the $M$ dyons, but $-1$ for ``additionally twisted"
$L$ dyons. 

Their
result is shown in the same figure by the filled diamond, and its suggests the $M$ density
  about twice larger than in the other method. Unfortunately, the latest study was 
done only at one temperature $T=1.5T_c$. The factor two discrepancy  between the results in Ref. \cite{Ilgenfritz:2006ju} and in Ref. \cite{Bornyakov:2008im} is 
due to different efficiency of their topology-searching algorithms.   The values of the the paper Ref.\cite{Ilgenfritz:2006ju}  mean to be the lower bound on the densities, as the authors also report also
 rather large number of  ``unidentified topological clusters" in the same table. 

It is instructive to compare the dyon densities observed on the lattice with the analytic semi classical expressions. Those include  the holonomy $\nu(T)=v(T)/(2\pi T)$ and the effective action $S_0=8\pi^2/g^2$ 
in a combination
\be  n_M=C \nu^{8\nu/3} \exp(-S_0 \nu),\,\,\,  \ee
where
$C$ is the normalization constant and the product of the pre-exponent powers of the holonomy is
taken from  the DGPS weight for the LM pair \cite{Diakonov:2004jn}.  The density of $L+\bar L$-dyons is given by obvious change $\nu\rightarrow \bar{\nu}=1-\nu$.Since the holonomy $\nu(T)$ goes to small
values at high $T$ more rapidly than the charge is running, one expects 
a significant excess of the $M$-dyons over the $L$-type ones at high $T$. The difference
between the two densities 
should  however disappear at $T=T_c$, with a characteristic singularity related to critical behavior of the SU(2) theory.

 As it is clear from  Fig.\ref{fig_dens} , the observed dependence shown by curves
can be well reproduced by the semiclassical expression, with two fitted parameters:
 \be C=1.6 fm^{-4}, S_0=3 
 \ee 
We also note that the point from \cite{Bornyakov:2008im} at $T=3/2Tc\approx \,450 \,\,\text{``MeV"}$ corresponds to
rather small dimensionlless density 
\be n_{L+\bar{L}}/T^4\approx 0.007. \ee 
or volumes/dyon above even the highest values used in our simulations.

We recall that the lattice data apply to a fully interacting ensemble, e.g. including the screening part of the action and 
 that the data/expressions discussed so far apply only to a pure gauge theory, i.e. to
$N_f=0$, which is characterized by the weakest coupling along the critical line. 
Hence,  by increasing number of quark flavors we 
expect to assess a significantly stronger coupling domain, hence a smaller $S_0\sim 1$. 
 If so,  eventually there would be no numerical suppression
 and  the diluteness should become $n_{L+\bar{L}}/T^4\sim O(1)$.  Such a dense liquid 
  of (non-semiclassical) dyons  is obviously not expected to be described by the simple formulas  used
above. This regime should be the subject of future theoretical and numerical studies.  

 While grouping of the dyons into instantons suggests about equal number of $L,M$ dyons,
this equality is only reached at and below $T_c$, while at $T>T_c$ the lighter $M$ dyons are more
numerous than heavier $L$ ones. The relevent lattice data are shown in Fig.\ref{fig_dens}.
As one can see, the density of $M$-dyons grows at high temperature, as their action decreases. The density of $L$-type,
associated with calorons, decreases, at large $T$. 
We emphasize again  that the identified topological objects in the lattice simulations are just a sub-set of the
of the observed topological clusters.

Only the solid diamond corresponding to the analysis of Bornyakov \emph{et al.} \cite{Bornyakov:2008im} should be considered as a quantitative
determination of the $L$-type dyon density. Indeed, this was  done by combining the fermionic quasizero-modes filtering procedure with the action smearing to locate topology, and by computing the  sign of the Polyakov line at the center to identify the $L$ and $M-$type dyons.
 Such a  work focused on $T = 1.5~T_c$ SU(2) gauge configurations in a  $20^3\times 4$ lattice. These authors confirmed 
that several main predictions of the semiclassical theory remain valid in their lattice configurations. 
 In particular, all fermionic zero and near-zero-modes are locally chiral ( a general property of their topological nature), the periodic fermions interact with $M$ and antiperiodic fermions with $L$ dyons, as expected from semiclassical solutions. The $L$ dyons are dilute and paired with $\bar{L}$ ones into Òclusters" discussed in \cite{Shuryak:2012aa} .  The chiral symmetry of the antiperiodic fermions is unbroken. The $M$-type dyon ensemble is dense and thus chiral symmetry for periodic fermions is broken. They also found that while
 the  L-type dyons occupy only  about 3\% of the lattice volume, the M dyons occupy a significant fraction of it.

\end{document}